%% file: mued-higgs.tex
\documentclass[aps,superscriptaddress,nofootinbib,A4,floatfix,preprint]{revtex4} 
\usepackage{graphicx,epsfig}
\usepackage{amsfonts}
\usepackage{amsmath}
\usepackage{amssymb}
\usepackage{array}
\usepackage{color}
\usepackage{hyperref}
\usepackage{longtable}
\usepackage{slashed}
\usepackage[novbox]{pdfsync}
\renewcommand{\d}{\ensuremath{\mathrm{d}}}
\newcommand{\ii}{\ensuremath{\mathrm{i}}}
\newcommand{\D}{\ensuremath{\mathcal{D}}}

\begin{document}

\preprint{UT-HET-071}

\title{Testing Minimal Universal Extra Dimensions Using Higgs Boson Searches at the LHC}

\author{Genevi\`eve B\'elanger}
  \affiliation{LAPTH, Universit\'e de Savoie, CNRS, B.P.110, F-74941
  Annecy-le-Vieux Cedex, France}

\author{Alexander Belyaev}
  \affiliation{School of Physics \& Astronomy, University of Southampton, UK}
  \affiliation{Particle Physics Department, Rutherford Appleton Laboratory, Chilton, Didcot, Oxon OX11 0QX, UK}
\author{Matthew Brown}
  \affiliation{School of Physics \& Astronomy, University of Southampton, UK}
\author{Mitsuru Kakizaki}
  \affiliation{Department of Physics, University of Toyama, 3190 Gofuku,
  Toyama 930-8555, Japan}
\author{Alexander Pukhov}
  \affiliation{Skobeltsyn Institute of Nuclear Physics, Moscow State
  University, Moscow 119992, Russia}

\begin{abstract}

Large Hadron Collider (LHC) searches for the SM Higgs boson provide a powerful limit on 
models involving Universal Extra Dimensions (UED) where the Higgs production is enhanced.
We have evaluated  all one-loop diagrams for Higgs production 
$gg\to h$ and decay $h\to \gamma\gamma$ within ``minimal'' UED (mUED), 
independently confirming previous results, 
and we have evaluated enhancement  factors for Higgs boson production and decay over the mUED parameter space.
Using these we   have derived limits on 
the parameter space, combining data from both  ATLAS and CMS collaborations
for the most recent 7 TeV  and 8 TeV LHC data.
We have performed a  rigorous statistical combination of several Higgs boson search channels which
is important because mUED signatures from the Higgs boson  are not universally enhanced.

We have found that $R^{-1}<500$~GeV is excluded at 95\% CL, while
for larger  $R^{-1}$ only a very narrow ($\pm 1-4$~GeV) mass window around
$m_h=125$~GeV and another window (up to 2~GeV wide for $R^{-1} > 1000$~GeV) around $m_h = 118$~GeV are left. The latter is likely to be excluded as more data becomes available whereas the region around $m_h = 125$~GeV is where the recently discovered Higgs-like particle was observed and therefore where the exclusion limit is weaker. 

It is worth stressing that mUED predicts 
an enhancement for all  channels for $gg\to h$ production and decay while  the vector boson fusion process $WW/ZZ \to h \to \gamma\gamma$
is generically suppressed  and  $WW/ZZ \to h \to WW^*/ZZ^*$ is standard. Therefore, 
as more 8 TeV LHC data becomes available, the information on individual Higgs  boson  production and decay processes  provided by the CMS and ATLAS experiments can be effectively used to  favour mUED or exclude it further.
\end{abstract}
\maketitle

\input{01-intro-mued-higgs}
\input{02-model-mued-higgs}
\input{03-calcs-mued-higgs}
\input{04-stats-mued-higgs}
\input{05-results-mued-higgs}

\input{06-conclude-mued-higgs}

\section*{Acknowledgements}

We would like to thank Aleksandr Azatov for useful discussion on the statistical methods.
The work of GB, AP was supported in part by the GDRI-ACPP of CNRS.
AB and MB thank  the Royal Society for  partial financial support. AB also
thanks  the NExT Institute and SEPnet for financial support.
The work of AP was supported by the Russian foundation for Basic Research, 
grant RFBR-10-02-01443-a. 
The work of MK was supported in part by
Toyama Prefecture Citizens' Personal Development Foundation (TPCPDF).

\bibliographystyle{h-physrev5}
\bibliography{mued-higgs}
\include{appendix-mued-higgs}

\end{document}

%% file: 01-intro-mued-higgs.tex
%!TEX root = /Users/matt/Documents/Extra dimensions/Higgs/dropbox_contents/higgs_paper/mued-higgs.tex

\section{Introduction}

Theories with Universal Extra Dimensions (UED) are very promising for solving
puzzles in the Standard Model (SM). The UED framework was proposed by
Appelquist \emph{et al} \cite{PhysRevD.64.035002}, following the suggestion of
the existence of large (i.e. millimetre-scale) extra dimensions
\cite{ArkaniHamed:1998rs,antoniadis1998new} or a warped (Planck-scale)
extra dimension~\cite{PhysRevLett.83.3370}. In UED, unlike in the preceding
extra dimension models, all SM particles are postulated to propagate in a
TeV$^{-1}$-sized bulk (normal space plus the extra compactified dimensions).
Models of UED provide solutions to problems such as explaining the three
fermion generations in terms of anomaly cancellation \cite{Dobrescu:2001ae},
and providing a mechanism for a sufficient suppression of proton
decay~\cite{Appelquist:2001mj}. Moreover, UED models can naturally incorporate
a $\mathbb{Z}_2$ symmetry called KK parity, analogous to $R$ parity in
supersymmetry, leading to a well-motivated dark matter candidate
\cite{Cheng:2002ej, Servant:2002aq}.

The simplest UED theory is known as minimal Universal Extra Dimensions
(mUED) and it posits a single, flat extra dimension compactified on an
$S^1/\mathbb{Z}_2$ orbifold in order to recover chiral interactions in
the 4D effective theory. Periodicity on a circle ($S^1$) leads to the
discretisation of momentum along the extra dimension into integer
multiples of the compactification scale, i.e. $p_5 = n R^{-1}$, where
$R$ is the radius of the circle. The integer $n$ is called
``Kaluza-Klein (KK) number'' and is a conserved quantity\footnote{
It is conserved in the following sense. Consider a vertex of three particles
with KK numbers $n$, $m$ and $l$. This vertex ``conserves'' KK number if $\pm n
\pm m \pm l = 0$ can be satisfied for some (independent) choice of plus or
minus signs. E.g. a $(1,1,0)$ vertex would conserve KK number, but $(0,0,1)$
would not.
}
before orbifolding. The ``orbifolding'' to $S^1/\mathbb{Z}_2$ leads to KK
number conservation being violated at loop level. However, KK \emph{parity} -- defined
to be $(-1)^n$ -- is conserved to \emph{all} orders in perturbation
theory. As a result of this symmetry, mUED predicts a stable lightest
Kaluza-Klein particle (LKP) which would be a prospective candidate for
dark matter.

The collider phenomenology of mUED has been studied intensively in many
publications (e.g.\cite{Macesanu:2002db, Bhattacherjee:2009uq,
Choudhury:2009kz, Bandyopadhyay:2009gd, Murayama:2011hj}), but we are only
aware\footnote{We thank Kohsaku Tobioka for bringing this work to our
attention.} of one \emph{experimental} paper \cite{ATLAS-CONF-2011-155} that
has set LHC limits on mUED. This is not surprising -- the search for mUED is
much more difficult than the search of SUSY within the experimentally
well-explored mSUGRA scenario at the LHC. The main reason for this is that mUED
provides much smaller missing transverse momentum due to the small mass
splitting between KK-partners of SM particles of the same KK level. Though dark
matter constraints set an upper limit on the scale of mUED below about
1.6~TeV~\cite{Belanger:2010xy}, this scale will be very difficult to test even
with the 14~TeV LHC~\cite{Murayama:2011hj, belyaev:2012a}. More pragmatically,
only a few computational tools for studying mUED are easily accessible to
experimentalists: Datta \emph{et al}~\cite{Datta:2010us} implemented mUED in
\texttt{CompHEP{\rm\cite{Pukhov:1999gg,Boos:2004kh}}/CalcHEP}~\cite{Pukhov:2004ca,Belyaev:2012qa} and independent implementations
\cite{Belanger:2010xy,belyaev:2012a} have improved upon this by treating
electroweak symmetry breaking consistently.

We show in this paper that the Higgs sector of mUED provides an excellent way
of testing the model at the LHC as was shown recently in
\cite{Nishiwaki:2011gk,Belanger:2012zg}. Indeed the loop-induced production process $gg\to h$ and decay $h\to \gamma\gamma$ are sensitive to heavy KK particles and are thus
different from their SM values. Here we improve on previous results by
rigorously combining the limits from different channels ($gg\to h \to
\gamma\gamma$, $gg\to h \to W^+ W^- \to \bar\ell\nu\ell\bar\nu$ and $gg \to h
\to ZZ \to 2\bar\ell 2\ell$) statistically, using the latest ATLAS and
CMS Higgs search results. Constraints on the mUED parameter space are then
derived. Going beyond \cite{Nishiwaki:2011gk}, we also show the effects of
including the radiative mass corrections for these particles. Our
independently-derived expressions for $gg\to h$ and $h\to\gamma\gamma$
amplitudes agree with those derived first by Petriello \cite{Petriello:2002uu}
%(which were used in \cite{Nishiwaki:2011gk}) when tree-level values of the KK masses are used.

This paper is structured as follows. The next section describes the main
features of the mUED model. In Sec.~\ref{sec:calcs}, we evaluate and present the effect
of KK-particles in the loop for $gg\to h$ production and Higgs decay to
$\gamma \gamma$, $W^+W^-$ and $ZZ$. The impact of KK-particles differ for
each channel and this non-universality should be taken into account when
establishing combined limits on the mUED parameter space. We express the
results of that section in terms of the enhancement of the $gg \to h \to
\gamma\gamma$, $gg \to h \to W^+ W^-$ and $gg \to h \to ZZ$
cross-sections. Next, in Sec.~\ref{sec:stats}, we discuss how these results
can be used to constrain the parameter space of the model, describing the
problems encountered when statistically combining experimental data from
different channels. In Sec.~\ref{sec:results} we show new limits on the
mUED parameter space using our rigorous statistical combination and the latest
ATLAS and CMS data. Section~\ref{sec:conclusions} contains our conclusions.
Details on the calculation of the $gg\to h$ and $h\to \gamma\gamma$ amplitudes
can be found in a set of appendices.

%% file: 02-model-mued-higgs.tex
%!TEX root = /Users/matt/Documents/Extra dimensions/Higgs/dropbox_contents/higgs_paper/mued-higgs.tex

\section{The mUED model}

In UED, in contrast to other Kaluza-Klein theories, there is one or more
towers of KK particles associated with \emph{every} SM particle. The particles
in a KK tower each have the same quantum numbers but progressively heavier
masses. In mUED, to a good approximation, the mass of a KK particle is given
in terms of its KK number by $n R^{-1}$, leading to a very regularly-spaced
mass spectrum. At the tree level, this regular spacing is altered slightly by
electroweak contributions $m_0$ to the mass so that
\begin{equation}
    m_n^{\text{tree}} = \sqrt{n^2/R^2 + m_0^2}\, . 
\end{equation}
Furthermore, radiative corrections to the KK masses play a crucial role.
Corrections to the masses of the strongly-interacting KK particles can be as
large as 30\% and, even for the weakly interacting particles for which the mass
corrections are numerically small, radiative effects are extremely important.
Without them there would be many nearly-degenerate particles and all the KK
partners of (nearly) massless SM particles would be stable to a good
approximation. Radiative corrections, first calculated in \cite{Cheng:2002iz},
lift the degeneracy. This means that all KK particles eventually decay to SM
particles and the lightest KK particle (LKP), which is forbidden to decay to SM
particles by KK parity conservation.  This LKP (a heavy version of the photon
for much of the parameter space) is an excellent dark matter candidate. The
small mass splittings between KK-partners of SM particles of the same KK level
leads to soft jets and leptons in the decay of KK-particles thus making it more
challenging to extract a signal at the LHC.

Associated with the SM $W^\pm$ boson there is a single tower of KK partners
$W_n^\pm$. However, each SM fermion $f$ has \emph{two} KK towers
denoted $f_{1,2}^{(n)}$. This feature will be relevant when comparing the size
of the contribution of bosons and fermions to the Higgs partial widths. Without
electroweak and radiative corrections, these particles have simple
interpretations: they are the KK partners of the $\text{SU(2)}_L$ doublet and
singlet respectively and only the left-handed (right-handed) components of
$f_1$ ($f_2$) survive at the zero KK level after the orbifold projection.

The KK modes, on the other hand, are vector-like, i.e. their left- and
right-handed components transform in the same way under SU(2). Another way to
say this is that both components couple equally to the KK $W^\pm$ bosons. With
electroweak and radiative effects included however, the mass eigenstates become
mixtures of the electroweak eigenstates and so the couplings to the gauge
bosons become chiral.

% \subsection{The mUED parameter space}
There are two free parameters in mUED: the Higgs mass $m_h$ and the
compactification scale $R^{-1}$. Strictly speaking, because mUED (like all
theories involving extra dimensions) is not renormalisable it must be treated
as an effective theory valid to some specified cut-off momentum scale
$\Lambda$. Thus $\Lambda$ is technically a third parameter of the theory. In
practice, however, low energy observables are only weakly sensitive to the
cut-off. For definiteness, in this paper, like in many of earlier works, as a
benchmark point we take $\Lambda = 20 R^{-1}$ which is low enough to keep the
SM coupling constants perturbative below the cut-off
scale~\cite{PhysRevD.64.035002,Bhattacharyya:2006ym}.\footnote{The vacuum stability condition constrains the cutoff scale $\Lambda R
    \lesssim 5$ for $R^{-1} \sim 1\text{ TeV}$ and $m_h = 125$ GeV
\cite{Blennow:2011tb}. This bound can be evaded if the SM vacuum is
metastable below the cutoff scale.}

In mUED the Higgs mass is limited to be below around 230 GeV by the simple requirement
that the dark matter candidate should be neutral \cite{Matsumoto:2005uh,
Belanger:2010xy}.
 More stringent limits are derived from collider searches for
the SM Higgs boson. Indeed as we will demonstrate in this paper, the signals
from the Higgs boson in mUED are enhanced as compared to those of the SM in
nearly all of the main search channels. One exception is the W-fusion
production of the Higgs decaying to two photons. The LEP limit on the SM
Higgs, $m_h > 114.4$ GeV, therefore provides a conservative lower limit.
The LHC sensitivity to the Higgs within the mUED scenario is better than for
the SM Higgs boson, leading to a reduced range of allowed masses as we will
derive in the next sections. 
As we know, recently the discovery of the Higgs-like particle with $m_h$=125 GeV was claimed by both the CMS~\cite{Chatrchyan:2012gu}
and ATLAS~\cite{Aad:2012gk} collaborations. This signal has a strong effect on the mUED parameter space and 
we use these latest CMS and ATLAS results (expressed in the form the limits on the SM Higgs parameter space)
to limit mUED with $m_h$ around 125~GeV.
By the end of 2012 the LHC will be
able to collect more statistics and clarify the nature of the Higgs-like particle
which eventually could be applied to further uncover the status of mUED.

A lower bound of around 600 GeV on the compactification scale comes from tests
of electroweak precision measurements \cite{Gogoladze:2006br} and $b \to s
\gamma$ \cite{Haisch:2007vb}. The upper bound on $R^{-1}$ is provided by
cosmological observations from the requirement that the abundance of the LKP
(whose mass is approximately $R^{-1}$) does not exceed the observed dark
matter abundance \cite{Belanger:2010xy}.

%% file: 03-calcs-mued-higgs.tex
%!TEX root = /Users/matt/Documents/Extra dimensions/Higgs/dropbox_contents/higgs_paper/mued-higgs.tex
\section{Evaluation of amplitudes for Higgs production and decay in mUED}
\label{sec:calcs}

\begin{figure}
  \begin{center}
    \includegraphics[width=0.23\textwidth]{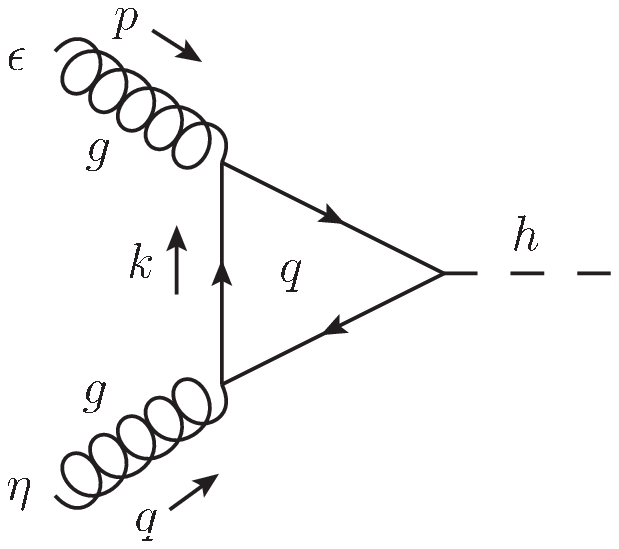}
    \hspace{5mm}
    \includegraphics[width=0.23\textwidth]{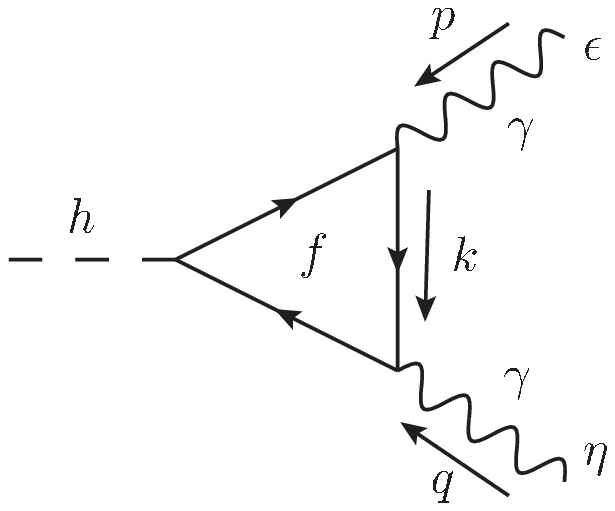}
    \includegraphics[width=0.23\textwidth]{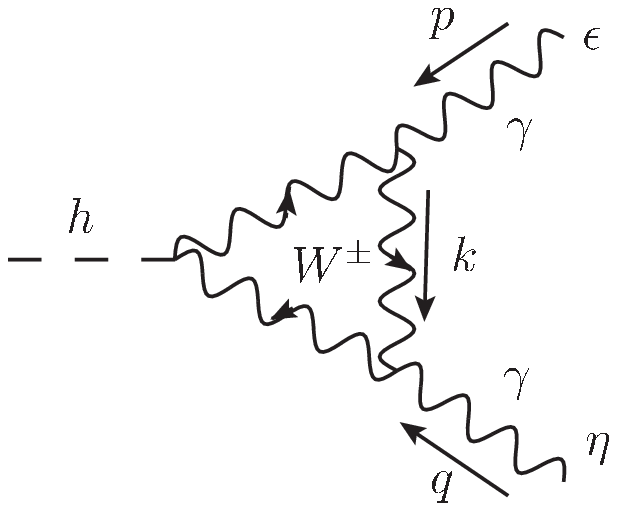}
    \includegraphics[width=0.23\textwidth]{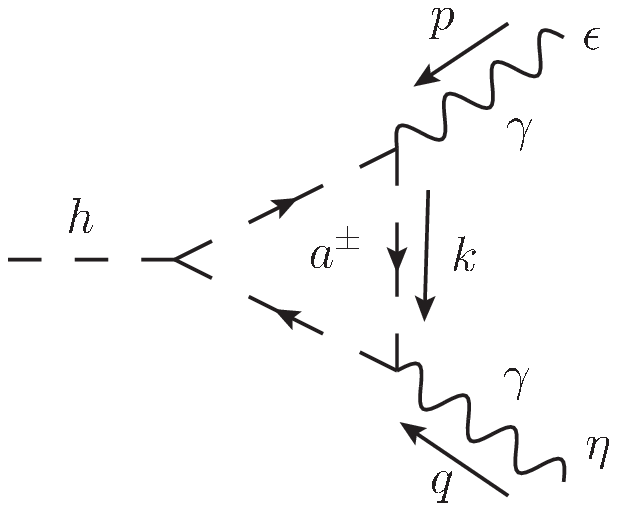}
  \end{center}
  \caption{Some diagrams involved in the production (left) and decay (right) of the
  SM Higgs boson. For the Higgs decay, there are also diagrams involving
  Goldstone bosons and Faddeev-Popov ghosts which are shown in
  appendix~\ref{app:haa}.}
\label{fig:feyndia}
\end{figure}

In the SM, the dominant process for producing the Higgs boson at the LHC is
gluon-gluon fusion, despite the leading order contribution being a one-loop
process. This process, shown in Fig.~\ref{fig:feyndia} (left), involves
triangle diagrams of quarks -- predominantly the top quark because of its large
Yukawa coupling. It is this large coupling and also the high gluon luminosity
at the LHC that makes this production mechanism dominant.  In mUED, KK quarks
can also run in the triangle loop leading to an enhancement over the SM
amplitude.

%gb
For low values of the Higgs mass (e.g. around the recently discovered \cite{Aad:2012gk,Chatrchyan:2012gu}
Higgs-like particle at 125~GeV) the most powerful Higgs search channel is into two photons. Indeed
the low QCD background for this process compensates for the fact that the Higgs
decay width into two-photons is  loop-induced and thus suppressed.  In the SM the
dominant contribution to the two-photon amplitude comes from loops involving
$W^\pm$ bosons. This contribution is about four times larger 
 than the one from
fermions. Furthermore, the charged fermion triangle loop (again, dominated by
top quarks) interferes destructively with the $W^\pm$ contribution.

%Both loop-induced production and decay processes are particularly sensitive to new physics such as mUED.
In mUED, new contributions arise from KK $W$'s and KK fermions running in
loops.  The contributions of the KK $W$'s and KK fermions have the same sign as
their SM counterparts, but the increase as compared to the SM contribution
is  larger for fermions than for W's.  First, associated with each SM fermion
there are two towers of KK fermions while there is only one for $W^\pm$.
Second, the contributions of particles from higher KK levels decrease more
slowly for fermions than for W's, as we will see in the next section.
Furthermore, for  KK number $n \ge 1$, there is an additional contribution
from  charged scalars $a^\pm_n$. This field is a mixture of the KK modes of the
5th component of the charged vector field and the charged component of the
Higgs field. At each KK level, the charged scalar contributes with the \emph{same}
sign as the fermion diagrams. The net effect is therefore to suppress the Higgs
to diphoton decay rate relative to the SM prediction.  The three (fermion,
$W^\pm$ and $a^\pm_n$) contributions are shown in Fig.~\ref{fig:feyndia}.
Additional diagrams involving $W^\pm$ Goldstones and ghosts are presented in
Appendix~\ref{app:haa}.

In the following subsections we show the results of calculating the amplitude
for production of a SM Higgs boson from gluon-gluon fusion, and also
the amplitude for subsequent decay to two photons. The amplitudes
$\mathcal{A}$ for the $gg\to h$ and $h\to \gamma\gamma$ processes both take
the form
\begin{equation}
\mathcal{A} =
    \tilde{\mathcal{A}} \left[(p\cdot q) (\epsilon \cdot \eta)  -
    (p\cdot \eta) (q\cdot \epsilon)\right],
    \label{eq:redamplitudeoffshell}
\end{equation}
where the external vector particles with momenta $p$ and $q$ have polarisation
vectors $\epsilon$ and $\eta$ respectively. These polarisation and momentum
conventions are shown in Fig.~\ref{fig:feyndia}.

In \eqref{eq:redamplitudeoffshell} the Higgs is allowed to be off-shell. To
calculate the exact amplitude for $gg \to h \to \gamma\gamma$, one would
combine the separate off-shell amplitudes for $gg \to h$ and $h \to
\gamma\gamma$ with a Higgs propagator. However, in our analysis we use the
``narrow width approximation'' (valid when the Higgs boson's width is much less
than its mass) which allows us to write the $gg \to h \to \gamma\gamma$ cross
section as the product of the cross-section for production of an on-shell Higgs
boson and the branching ratio of an on-shell Higgs to two photons, i.e.
\[
    \sigma(gg \to h \to \gamma\gamma) \approx \sigma(gg \to h) \times 
    \text{BR}(h \to \gamma\gamma).
\]

In this approximation, we only need amplitudes involving on-shell Higgs bosons,
so we can write \eqref{eq:redamplitudeoffshell} as
\begin{equation}
\mathcal{A} =
\tilde{\mathcal{A}} \left[\frac{m_h^2}{2} (\epsilon \cdot \eta)  -
    (p\cdot \eta) (q\cdot \epsilon)\right],
    \label{eq:redamplitude}
\end{equation}
where $m_h$ is the Higgs mass.
    
These amplitudes have been calculated previously in the SM case in the $m_h/M
\ll 1$ limit \cite{Ellis:1975ap} (where $M$ is the mass of the particle flowing
in the loop) and subsequently~\cite{Shifman:1979eb} for general $m_h/M$. They
have also been calculated in the mUED case (without radiative mass corrections)
in \cite{Petriello:2002uu}. We performed the calculation in the general mass
case for mUED and included radiative corrections to the KK masses for the
first time. We used the 't Hooft-Feynman gauge and regulated the divergences
that appear in intermediate steps using dimensional regularisation. We made use
of the well-known Passarino-Veltman functions~\cite{tHooft:1978xw} to evaluate
the momentum integrals. Our calculation is shown in detail with all
contributing diagrams in Appendices~\ref{app:ggh} and \ref{app:haa}. Our
results reduce to the SM result found in the literature~\cite{Djouadi:2005gi}
when the KK modes are removed and agree with the result in
\cite{Petriello:2002uu} when we use tree-level KK masses in mUED.

\subsection{Higgs production}
The amplitude for  $gg\to h$ (Fig.~\ref{fig:feyndia}) reads\footnote{It 
should be noted that higher loop corrections to the $gg\to h$ amplitude
can be substantial, reaching as much as  90\% of the one-loop amplitude
\cite{Spira:1997dg}. 
However, these large corrections are dominated by SM
contributions (KK contributions are suppressed by powers of the
compactification scale $R^{-1}$). The SM QCD corrections
depend only on spin of the particle in the large mass limit
and therefore they are universal for SM and mUED.
In this paper we are ultimately  only
interested in the ratio of mUED and SM rates in which the QCD corrections
cancel to a good approximation and therefore our results are valid for higher order QCD
corrections.
}

\begin{equation}
\tilde{\mathcal{A}}_{ggh} = -\frac{\alpha_{s}}{4\pi v}\left(F_{ggh}^{\text{SM}} + \sum_{n=1}^N F_{ggh}^{(n)}\right).
\label{eq:gghmaster}
\end{equation}
where $\alpha_s$ is the strong coupling constant and $v=2\sin\theta_W m_W / e$
is the Higgs vacuum expectation value ($\theta_W$ is the Weinberg angle, $e$ is
the elementary electric charge and $m_W$ is the mass of the $W$ boson).
In the SM there would be contributions from each quark flavour $q$  in
the loop, such that  $F_{ggh}^{\text{SM}} = \sum_{q} f_{F}(m_q)$ where the standard
fermion contribution is given by
\begin{equation}
f_{F}(m) = \frac{8m^{2}}{m_{h}^{2}}\left[1 + \left(1-\frac{4m^{2}}{m_{h}^{2}}\right)c_{0}(m)\right],
\label{eq:fF}
\end{equation}
where $c_{0}$ is a dimensionless form of the scalar three-point Passarino-Veltman
function
\[ c_{0}(m) = 
\begin{cases}
	\left[\arcsin\left(\frac{m_{h}}{2m}\right)\right]^{2} & m^{2} \ge m_{h}^{2}/4 \\
	-\frac{1}{4}\left[\ln\left(\frac{1+\sqrt{1-4m^{2}/m_{h}^{2}}}{1-\sqrt{1-4m^{2}/m_{h}^{2}}}\right) - i\pi\right]^{2} & m^{2} < m_{h}^{2}/4. 
\end{cases}
\]
Note that $f_F$ and $c_0$ are dimensionless functions and the argument $m$ always appears in the dimensionless combination $4m^2/m_h^2$ (often written as $\tau$ in the literature).

In the $m\gg m_h$ limit, the above expressions reduce to 
$c_0 = \frac{m_h^2}{4m^2} + \frac{m_h^4}{48m^4} + \mathcal{O}(m_h^6/m^6)$ and
\begin{equation}
    f_F(m) \approx \frac{4}{3} + \frac{m_h^2}{6m^2} + \mathcal{O}(m_h^4/m^4) \to 4/3.
\label{eq:fFasymp}
\end{equation}
Thus the amplitude tends to a constant in the heavy quark limit. This is however not the case when KK quarks are included in the loop:
the heavy KK quarks ``decouple''
and, therefore, progressively higher KK modes lead to progressively smaller modifications to the Higgs boson coupling. Consequently, as we show below, one can safely neglect higher KK modes.

The reason for this decoupling is that, in contrast with SM fermions,   
while a KK particle's mass increases with KK
number there is \emph{no} corresponding increase in its Yukawa couplings and
so decoupling does  occur because of suppression from the propagators. This decoupling behaviour is shown explicitly below.

In mUED, the contribution from KK quarks at the $n$th KK level (there are two KK quarks at each level for each SM quark $q$) is
\begin{equation}
F_{ggh}^{(n)} = \sum_{q}\sin(2a_{q}^{(n)})
\left(\frac{m_q}{m_{q,1}^{(n)}} f_{F}(m_{q,1}^{(n)}) +
\frac{m_q}{m_{q,2}^{(n)}} f_{F}(m_{q,2}^{(n)}) \right).
\label{eq:fgghfull}
\end{equation}
where $m_{q,1}^{(n)}$ and $m_{q,2}^{(n)}$ denote the KK quark masses and
$a_{q}^{(n)}$ denote the mixing angles required to diagonalise the KK quark
mass matrices. At tree level, all KK quark  masses are nearly degenerate, 
\[
m_{q,\text{tree}}^{(n)} = \sqrt{m_{q}^{2}+ \frac{n^{2}}{R^{2}}},
\]
where $m_q$ is the  zero mode mass. Radiative corrections induce mass splittings between the KK fermions 
(see e.g. \cite{Cheng:2002iz}). Similarly, the mixing angles are
\[
a_{q,\text{tree}}^{(n)} = \frac{1}{2}\arctan\left(\frac{m_{q}R}{n}\right)
\]
at tree-level (so $\sin(2a_{q,\text{tree}}^{(n)}) = m_q / m_{q, \text{tree}}^{(n)}$), but radiative corrections alter this expression (see for
example \cite{belyaev:2012a}).

In our analysis we used one-loop corrected expressions for all masses and
mixings as detailed in \cite{Cheng:2002iz}, but it is illustrative to neglect
loop corrections and study the behaviour (just considering the top-quark
contribution, which is dominant) for that case that $m_t^{(n)} > m_t > m_h$:
\begin{equation}
  F_{ggh}^{(n)}  \approx 2\left(\frac{m_t}{m_t^{(n)}}\right)^2 f_F(m_t^{(n)}) \approx 2 \left(\frac{m_t}{m_t^{(n)}}\right)^2 \times \frac{4}{3},
  \label{eq:Fgghnasymp}
\end{equation}
throwing away terms in $m_h/m_t$ and $m_t R$ of order 3 or higher. This demonstrates the fact, mentioned above, that (in contrast to SM quarks) heavy KK quarks decouple from the process.

Taking the mass of the $n$th KK quark to be approximately $n/R$ and considering just the top quark, the total KK contribution to the amplitude is approximately
\begin{equation}
F_{ggh}^{\text{KK}} \equiv \sum_{n=1}^{N}F_{ggh}^{(n)} \approx 2 \times \frac{4}{3} m_t^2 R^2 \sum_{n=1}^{N} \frac{1}{n^2}.
\end{equation}
The sum is convergent as $N\to \infty$, thanks to the decoupling of the heavy KK particles. In this limit, $F_{ggh}^{\text{KK}} \to 4(\pi m_t R)^2/9$. So the momentum cutoff uncertainty is quite mild if one chooses a reasonably large value for it.
%gb I remove this because it is repeated below 
%In this paper we take the cutoff in the fifth dimension to be $20 R^{-1}$,  if one were to remove the cutoff, the result would be a fractional error in the KK contribution of around $1/20 = 5\%$.

The sum over KK modes $n$ is taken up to a cutoff $N$, corresponding to a
momentum cutoff in the extra dimension of $N R^{-1}$. Mild cutoff-dependence
is expected in perturbatively non-renormalisable theories such as mUED. In our
quantitative analysis we chose $N=20$ and included only $t$ and $b$ in the sum
over quark flavours $q$, which is an excellent approximation due to the size
of their Yukawa couplings compared to those of the lighter quarks. One should
note that for large $N$ the rest of the sum is proportional to $1/N$.
Therefore, for  $N=20$ our result is given with about 5\% accuracy as
compared to the full sum.

\subsection{Higgs decay to two photons}

The $h\to \gamma\gamma$ amplitude is given by
\begin{equation}
\tilde{\mathcal{A}}_{h\gamma\gamma} = - \frac{\alpha}{2\pi v} F_{h\gamma\gamma},
\label{eq:amphaa}
\end{equation}
where $\alpha$ is the fine structure constant, $v$ is the Higgs vacuum
expectation value (defined just below Eq.~\ref{eq:gghmaster}), and
\begin{equation} 
F_{h\gamma\gamma} = F_{h\gamma\gamma}^{\text{SM}} +
\sum_{n=1}^{N}F_{h\gamma\gamma}^{(n)}
\end{equation}
The SM part consists of a contribution from the $W^\pm$ vector bosons and fermions:
\begin{equation}
F_{h\gamma\gamma}^{\text{SM}} = f_{V}(m_{W}) +
\sum_{f}n_{c}Q_{f}^{2}f_{F}(m_{f}).
\label{eq:FhAASM}
\end{equation}
The sum  is taken over all SM fermions $f$, each with charge
$Q_{f}e$, setting $n_{c}$ to 3 for quarks and 1 for leptons.
The fermion loop function $f_{F}$ is the same as for the $gg\to h$ case, given in \eqref{eq:fF}, and the vector function $f_V$ (representing the $W^\pm$ and related Goldstone and ghost contributions) is
\begin{equation}
f_{V}(m) = -2 -12\frac{m^{2}}{m_{h}^{2}} -
24\frac{m^{2}}{m_{h}^{2}}\left(1-\frac{2m^{2}}{m_h^2}\right)c_{0}(m).
\label{eq:fV}
\end{equation}
In the large mass limit this tends to a constant
\begin{equation}
    f_V \approx - 7 - \frac{m_h^2}{2m^2} + \mathcal{O}(m_h^4/m^4) \to -7,
\end{equation}
showing that particles whose masses are proportional to their Yukawa couplings
do not decouple from the process, just as we saw in \eqref{eq:fFasymp} for the
production amplitude.

At the $n$th KK level the amplitude receives contributions from KK charged
fermions (two KK partners for each SM fermion) and the KK $W_{n}^\pm$ vector
boson. There is also a contribution from the charged scalar $a_n^\pm$ that
is not present at the SM level, so
\begin{equation}
	F_{h\gamma\gamma}^{(n)} = f_F^{(n)} + f_V^{(n)} + f_S^{(n)}.
\label{eq:fhgg:KK}
\end{equation}
The fermion contribution is the same as the quark contribution \eqref{eq:fgghfull} was for the Higgs production amplitude, up to colour and charge factors:
\begin{equation}
    f_F^{(n)} = \sum_{f}n_{c}Q_{f}^{2}\sin (2a_{f}^{(n)})
    \left(\frac{m_{f}}{m_{f,1}^{(n)}} f_{F}(m_{f,1}^{(n)}) +
    \frac{m_{f}}{m_{f,2}^{(n)}} f_{F}(m_{f,2}^{(n)})\right).
\end{equation}    
and so has a similar asymptotic behaviour to the one shown in
\eqref{eq:Fgghnasymp}. The sum over KK modes is therefore convergent as well.

The vector contribution is given in terms of the SM expression as follows and also decouples as $m_{W,n} \to \infty$, in contrast to the SM case:
\begin{equation}
    f_V^{(n)} = \frac{m_W^2}{m_{W,n}^2}f_V(m_{W,n}) \approx -\frac{7m_W^2}{m_{W,n}^2} + \mathcal{O}(m_W^4/m_{W,n}^4) 
    \to 0.
\end{equation}

The scalar contribution is given by
\begin{align}
  f_{S}^{(n)}(m_{a,n},m_{W,n}) &=
  \left[\frac{2m_{W}^{2}}{m_{W,n}^{2}}\left(1-\frac{2m_{a,n}^{2}}{m_{h}^{2}}
  \right) - 2\right] \left[1-\frac{4m_{a,n}^2}{m_h^2}c_{0}(m_{a,n}) \right].
\label{eq:fS}
\end{align}
At tree-level, $m_{a,n} = m_{W,n}$ so, keeping $m_W$ and $m_h$ constant, as we increase the KK scalar's mass,
\begin{equation}
    f_S^{(n)} \approx \frac{m_W^2}{m_{a,n}^2}\left(\frac{1}{3} +
    \frac{m_h^2}{6m_W^2}\right) + \mathcal{O}(m_W^4/m_{a,n}^4) \to 0,
\end{equation}
again demonstrating decoupling behaviour in the large KK mass limit.

In the SM case we can use the limits when the mass of the particle flowing in
the loop is large compared to the Higgs mass to estimate the relative
contributions from fermions and vectors, noting that they have opposite signs.
Including the charge and colour factors for the fermion case and considering
only the top quark, the ratio is $|f_V|/|n_c Q_t^2 f_F| \approx 7/\frac{16}{9}
\approx3.9$. Following the same procedure for contributions from level $n$ KK
particles (taking their masses to be approximately $n/R$) we find not only that
the vector and fermion contributions each have the same sign as their SM
counterparts but also that the ratio of vector to fermion contributions is
smaller than in the SM: recognising that there are \emph{two} KK top quarks,
$|f_V^{(n)}|/|f_F^{(n)}| \approx 3.9(m_W^2 / 2 m_t^2) \approx 0.42$, i.e. less
than 1. 
This suggests that the net effect of KK particles will be from the top
quark contribution which will thus interfere destructively with the SM
contribution from $W$'s, reducing the overall amplitude. In addition, there is
the charged scalar contribution which has the same sign as the fermion
contribution, reducing the amplitude further. This indication of amplitude
suppression is confirmed by the full calculation.

\begin{figure}
	\begin{center}
			\includegraphics{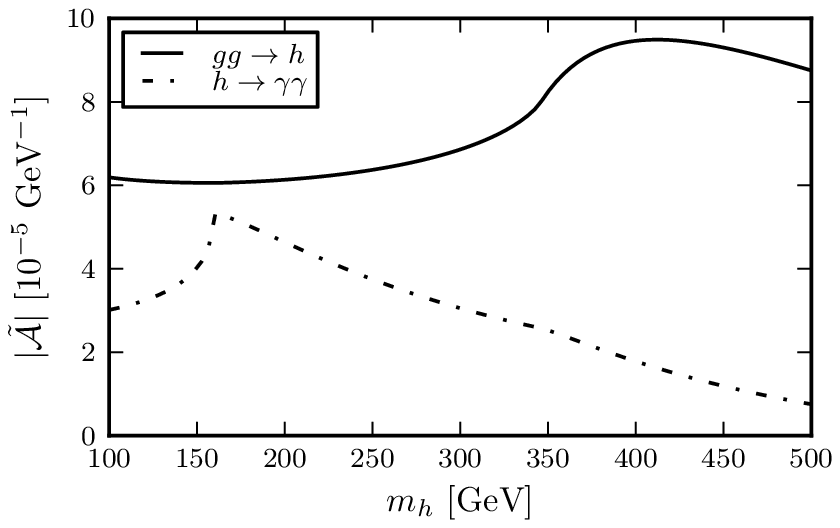}\\
			\includegraphics{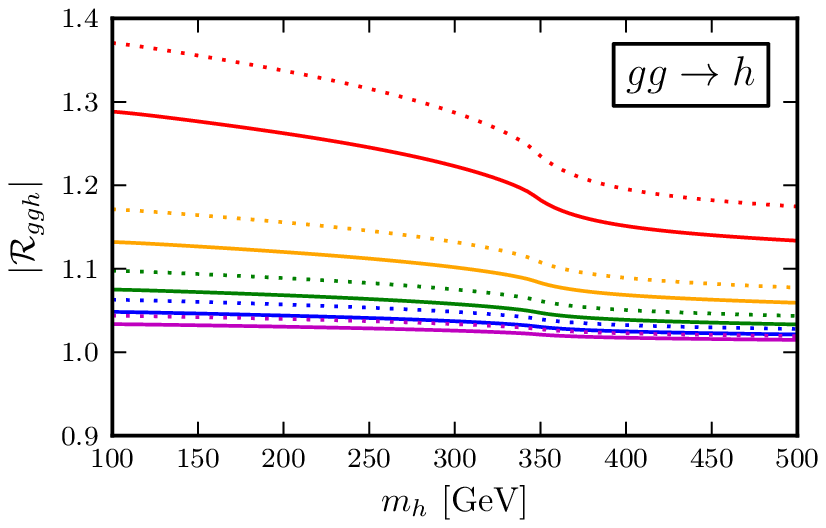}
			\includegraphics{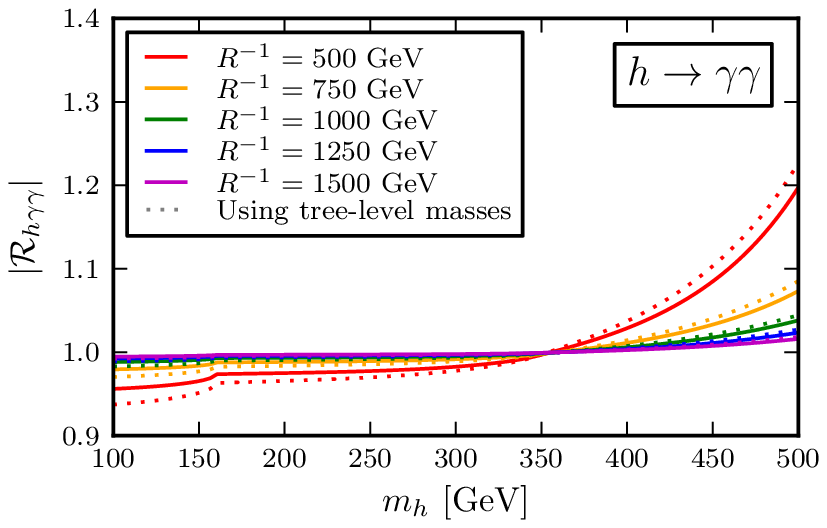}
	\end{center}
	\caption{Behaviour of the SM amplitudes and the relative sizes of the
corresponding mUED amplitudes for several values of $R^{-1}$. The top figure
shows the behaviour of the absolute values of the SM amplitudes for Higgs
production and decay to two photons respectively. The bottom figures show the
enhancement of these amplitudes in mUED relative to the SM, where
$\mathcal{R}=\mathcal{A}_{\text{UED}}/\mathcal{A}_{\text{SM}}$. For the mUED
plots, from top to bottom on the RHS of each plot: $R^{-1}=$ 500, 750, 1000,
1250 and 1500 GeV. Solid lines show the results when using loop-corrected KK masses and dashed lines show tree-level results.}
    \label{fig:mattratios}
\end{figure}

The dependence of the two amplitudes \eqref{eq:gghmaster} and
\eqref{eq:amphaa} on the two free parameters of mUED -- $m_{h}$ and the inverse
compactification radius $R^{-1}$ -- is shown in Fig.~\ref{fig:mattratios}.
This clearly indicates that for a light Higgs the $ggh$ coupling is enhanced
while $h\gamma\gamma$ is suppressed as argued above. The $R^{-1}$ dependence
enters through the KK masses and mixing angles. We have calculated the
amplitudes using tree-level KK masses (dashed lines) and loop-corrected values
(solid lines).

\subsection{Calculating the mUED cross-section enhancement}
\label{sec:cs_rat}
\begin{figure}
    \begin{center}
            \includegraphics{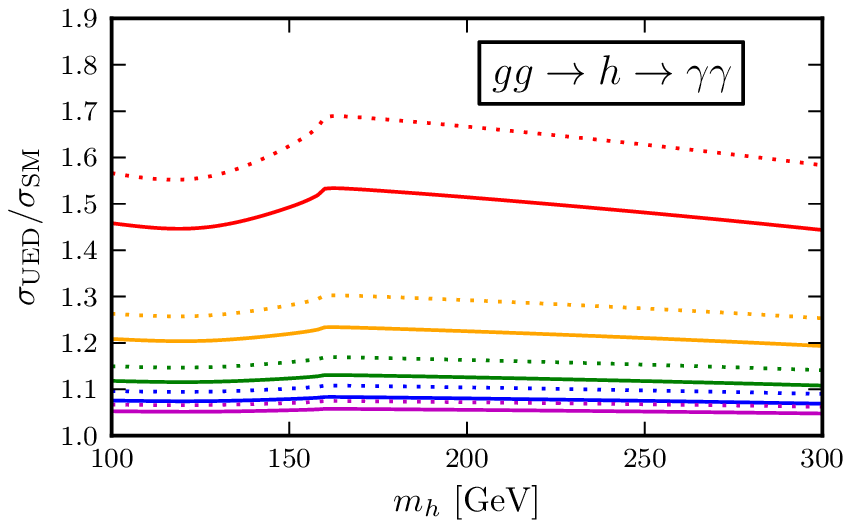}
            \includegraphics{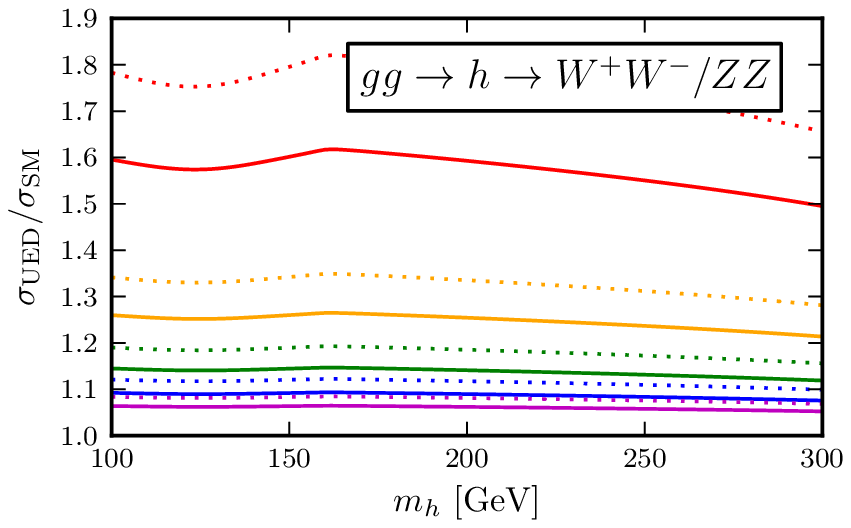}
    \end{center}
    \caption{Enhancement of mUED cross-sections in $\gamma\gamma$ (left) and $W^+
    W^-/ZZ$ (right) channels relative to the SM. The graphs show variation 
    with $m_h$ for the following values of $R^{-1}$: from top to bottom on 
    the RHS of each plot, $R^{-1}=$ 500, 750, 1000, 1250 and 1500 GeV. Solid lines show 
    results when using loop-corrected masses in the loops, while dashed lines 
    correspond to tree-level masses.}
    \label{fig:cs_rat}
\end{figure}

In order to constrain mUED using SM Higgs boson searches at the LHC, we first need
to calculate the enhancement of cross-sections of Higgs production and
subsequent decay in different channels. Here we consider the three most important
channels in the low Higgs mass range: $gg \to h \to \gamma\gamma$, $gg \to
h \to W^+W^- \to \bar{\ell}\nu \ell \bar{\nu}$ and $gg \to h \to ZZ \to 2\bar\ell 2\ell$. We can work in the narrow
width approximation $\Gamma_h \ll m_h$, assuming  that the Higgs is produced approximately
on-shell and subsequently decays with some branching ratio BR, so that
\[
    \sigma(xx \to h \to yy) = \sigma(xx\to h) \times\text{BR}(h\to
    yy).
\]
In fact, since we need only   the enhancement of each signal cross-section
relative to the SM, knowledge of the full hadronic cross-section is not required because the integrals of parton density functions would be the
same in mUED and the SM and would cancel in the ratio. The ratio can then be written simply (see, for example, \cite{Belyaev:2005ct}) in terms of total and partial Higgs widths as
\begin{equation}\label{eq:muaa}
    \mu_{\gamma\gamma} \equiv
    \frac{\sigma_{\text{mUED}}(gg \to h \to
    \gamma\gamma)}{\sigma_{\text{SM}}(gg \to h \to \gamma\gamma )}
    \approx
    \frac{\Gamma_{\text{mUED}}(h\to gg) \times \text{BR}_{\text{mUED}}(h \to
    \gamma\gamma)} {\Gamma_{\text{SM}}(h \to gg) \times
    \text{BR}_{\text{SM}}(h\to \gamma\gamma)}
\end{equation}
for the diphoton channel and
\begin{equation}\label{eq:muwwzz}
    \begin{split}
    \mu_{WW/ZZ} &\equiv
    \frac{\sigma_{\text{mUED}}(gg \to h \to
    WW)}{\sigma_{\text{SM}}(gg \to h \to WW)}
    \approx
    \frac{\Gamma_{\text{mUED}}(h\to gg) \times \text{BR}_{\text{mUED}}(h \to
    WW)} {\Gamma_{\text{SM}}(h \to gg) \times
    \text{BR}_{\text{SM}}(h\to WW)} \\
    &\approx
    \frac{\Gamma_{\text{mUED}}(h\to gg) \times \Gamma_{\text{SM}}(h\to
    \text{all})}
     {\Gamma_{\text{SM}}(h \to gg) \times
    \Gamma_{\text{mUED}}(h\to\text{all})}
    \end{split}
\end{equation}
for the $W^+ W^-$ and $ZZ$ channels. Note that the mUED and SM expressions for the partial Higgs
width to two vector bosons are the same, to leading order.

These two enhancement factors are plotted for various values of $m_h$ and
$R^{-1}$ in Fig.~\ref{fig:cs_rat}, also showing the effect of including loop
corrected masses in the loop diagrams.

%% file: 04-stats-mued-higgs.tex
%!TEX root = /Users/Matt/Dropbox/MUED-LAPTH/higgs_paper_response_to_comments/mued-higgs.tex
%!TEX root = /Users/matt/Documents/Extra dimensions/Higgs/dropbox_contents/higgs_paper/mued-higgs.tex

\section{Constraining the parameter space}
\subsection{Using one channel}
Results for experimental searches for the Higgs boson at the LHC (by the ATLAS and CMS collaborations) are usually presented using ``Brazil band'' combined
plots. These plots can be applied to family of models related to the SM in the
following way. The pattern of fully-exclusive Higgs signal cross-sections
($\sigma(xx \to h \to yy)$) is the same as the Standard Model's 
except that each of them is  scaled by some uniform factor, often denoted by $\mu$. The
plots show the value of this enhancement factor that is excluded 
at the 95\% confidence level for each value of the Higgs mass. This quantity
is normally written as $\mu^{95\%}$. When $\mu^{95\%}$ drops below unity, the
SM is excluded at the 95\% confidence level.

Although  $\mu^{95\%}$ can be used to exclude models that have the same pattern of cross-sections as the SM, for models (such as mUED) where different channels receive different corrections from new physics, this combined $\mu^{95\%}$ is not a useful quantity.
Fortunately, the collaborations also provide exclusion plots for separate channels. It is then a simple matter to compare the value of, say,  $\mu_{\gamma\gamma}$ to the excluded value $\mu_{\gamma\gamma}^{95\%}$. The exclusions from each channel and each experiment can then be overlapped in a simple way to constrain the model. As mentioned in the introduction, this has been done previously for mUED \cite{Nishiwaki:2011gk, Belanger:2012zg}.
However more accurate constraints on the model's parameter space can be obtained with  a more sophisticated 
method of combining the exclusions from different channels in a statistically rigorous way. Such a method is discussed in the next section.

\subsection{Statistical combination}
\label{sec:stats}
We want to reproduce as closely as possible the analysis used by the experimental
collaborations to calculate $\mu^{95\%}$ for the SM Higgs, but within the framework of mUED. We
start completely analogously by imagining a family of models, each exactly the
same as mUED except that the Higgs signal cross-sections in each channel are
all scaled by a common factor $\mu$. So, for example, if mUED (for certain
values of $m_h$ and $R^{-1}$) predicts a $gg\to h \to \gamma\gamma$
cross-section of $\sigma_{\gamma\gamma}^{\text{mUED}}$, a $gg\to h \to
WW \to \bar\ell\ell\bar\nu\nu$ cross-section of
$\sigma_{WW}^{\text{mUED}}$, and a $gg\to h \to
ZZ \to \bar2\ell 2\ell$ cross-section of
$\sigma_{ZZ}^{\text{mUED}}$, we imagine a family of related models
predicting $\{\mu\sigma_{\gamma\gamma}^{\text{mUED}},
\mu\sigma_{WW}^{\text{mUED}}, \mu\sigma_{ZZ}^{\text{mUED}}\} = \{\mu\mu_{\gamma\gamma}
\sigma_{\gamma\gamma}^{\text{SM}},
\mu\mu_{WW}\sigma_{WW}^{\text{SM}}, \mu\mu_{ZZ}\sigma_{ZZ}^{\text{SM}}\}$, writing the cross-sections in terms
of the mUED enhancement factors defined in \eqref{eq:muaa} and \eqref{eq:muwwzz}.

We then construct functions giving the probability of observing a particular
numbers of events in each channel (the ``individual likelihoods'', $p_i \equiv p(n_i^{\text{obs}}|\mu, \mu_i)$). These will
depend on the expected number of events in each channel $i$, given by
\[
n_i = s_i + b_i = \mathcal{L}\varepsilon_i \mu\mu_i\sigma_i^{\text{SM}} + b_i.
\]
Here, $s_i$ and $b_i$ denote the total number of signal and background events in
channel $i$ expected to be observed in the model defined by $(m_h, R^{-1}, \mu)$.
 The integrated luminosity is given by $\mathcal{L}$
and the signal cross-section can be written as $\mu\mu_i
\sigma_i^{\text{SM}}$. Finally, it should be noted that the number of
events one is able to see differs from the number of events that occur because
of detector inefficiencies, particle misidentification and kinematical cuts.
This is taken into account by the ``efficiency'' factor $\varepsilon_i$.

Once the individual likelihoods $p_i = p(n_i^{\text{obs}}|\mu, \mu_i)$ are
known, the total joint likelihood $P(\{n_i^{\text{obs}}\}|\mu, \{\mu_i\}) =
\prod_i p_i$ can be easily formed and then $\mu^{95\%}$ can be calculated.

The difficulty comes in reconstructing the likelihoods. The experimental
collaborations do not routinely make available the efficiency factors, exact
number of observed events after cuts, or expected number of background events
after cuts. What they \emph{do} make available is the value of $\mu_i^{95\%}$
for many of the channels, and also the ``expected''
$\mu_{i,\,\text{expected}}^{95\%}$, which is the probability that the number
of observed events might fluctuate down to the background-only expectation.

Azatov \emph{et al} propose \cite{Azatov:2012bz} a method for approximately
reconstructing the individual channel likelihoods from the data provided by
the experimental collaborations. We have followed their method in this paper,
and explain some important points here.

It is possible to write the likelihood approximately as
\[
    p_i \propto \exp\left[-\frac{(n_i^{\text{obs}} -
    n_i)^2}{2n_i^{\text{obs}}}\right]
    \propto \exp\left[-\frac{(\mu \mu_i -
    \beta_i)^2}{2\alpha_i^2}\right],
\]
 when $n_i^{\text{obs}} \gg 1$ (in fact
$n_i^{\text{obs}} > 10$ is a good approximation). Here we have introduced the
following quantities:
\[
    \alpha_i \equiv
    \frac{\sqrt{n_i^{\text{obs}}}}{s_i^{\text{SM}}}
    \quad\text{and}\quad
    \beta_i \equiv \frac{n_i^{\text{obs}}-b_i}{s_i^{\text{SM}}},
    \quad\text{where}\quad
    s_i^{\text{SM}} = \mathcal{L}\varepsilon_i \sigma_i^{\text{SM}}.
\]
The important point to realise is that we have managed to write the three
unknown quantities $n_i^{\text{obs}}$, $b_i$ and $\varepsilon_i$ in just two
independent combinations, $\alpha_i$ and $\beta_i$.

Making the further reasonable approximation that $(n_i^{obs} - b_i)/b_i \ll 1$
we can deduce, as shown in eq.~3.24 in \cite{Azatov:2012bz}, that
\[
    \alpha_i \approx \frac{\sqrt{b_i}}{s_i^{\text{SM}}} = \frac{\mu_{i,\,
    \text{expected}}^{95\%}}{1.96}
\]
if we interpret exclusion limits in the Bayesian sense. With this knowledge
we can then infer the value of $\beta_i$ from the \emph{observed} $\mu_i^{95\%}$,
provided by the experimental collaborations, by solving the following equation
(eq.~3.22 in \cite{Azatov:2012bz}) numerically:
\[
    0.95 \approx 
    \frac{\text{Erf}\left(\frac{\mu_i^{95\%} - \beta_i}{\sqrt{2}
    \alpha_i}\right)
    +
    \text{Erf}\left(\frac{\beta_i}{\sqrt{2}\alpha_i}\right)}
    {1 + \text{Erf}\left(\frac{\beta_i}{\sqrt{2}\alpha_i}\right)},
\]
where the error function $\text{Erf}(x) = \frac{2}{\sqrt\pi}\int_0^x \mathrm{e}^{-t^2}\mathrm{d} t$.

With the individual likelihoods approximately reconstructed in this way we
can form the joint likelihood and calculate the combined $\mu^{95\%}$ (again,
working in the Bayesian picture). We find it to be
\[
    \mu^{95\%} = \beta_{\text{comb}} + \sqrt{2}\alpha_{\text{comb}} \times
    \text{Erf}^{-1} \left[0.95 - 0.05 \times
    \text{Erf}\left(\frac{\beta_{\text{comb}}}
    {\sqrt{2}\alpha_{\text{comb}}}\right)\right],
\]
where
\[
    \alpha_{\text{comb}} \equiv \left(\sum_i
    \frac{\mu_i^2}{\alpha_i^2}\right)^{-\frac{1}{2}}
\]
and
\[
    \beta_{\text{comb}} =
    \alpha_{\text{comb}}^2 \times \sum_i\frac{\mu_i \beta_i}{\alpha_i^2}.
\]

Using the procedure outlined above, we performed a scan over the mUED
parameter space, calculating $\mu^{95\%}$ for each point $(m_h,R^{-1})$. We
used the $gg \to h \to \gamma\gamma$, $gg \to h \to W^+ W^- \to
\bar\ell\ell\bar\nu\nu$ and $gg \to h \to ZZ \to 2\bar\ell 2\ell$ channels
from ATLAS and
CMS Higgs boson searches. We scanned $m_h$ in 2-GeV
steps, $R^{-1}$ in 12.5-GeV steps. We have further considered additional
constraints on the parameter space. The Higgs mass range is bound from below
by LEP limits and from above by the requirement that the dark matter candidate
be neutral -- see~\cite{Belanger:2010xy}. The inverse radius must be greater
than around 600~GeV so as not to conflict with electroweak precision tests \cite{Gogoladze:2006br, Haisch:2007vb},
and less than 1600 GeV so that the dark matter candidate is not too heavy
\cite{Belanger:2010xy}.

%% file: 05-results-mued-higgs.tex
%!TEX root = /Users/matt/Documents/Extra dimensions/Higgs/dropbox_contents/higgs_paper/mued-higgs.tex
\section{Results}
\label{sec:results}

\begin{figure}
	\begin{center}
		\includegraphics{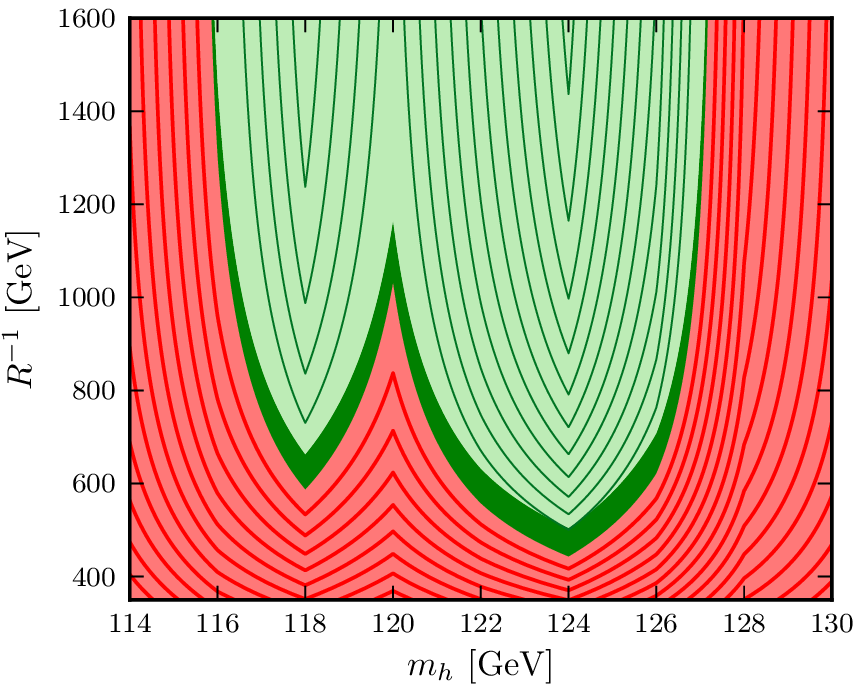}
        \includegraphics[width=0.5\textwidth]{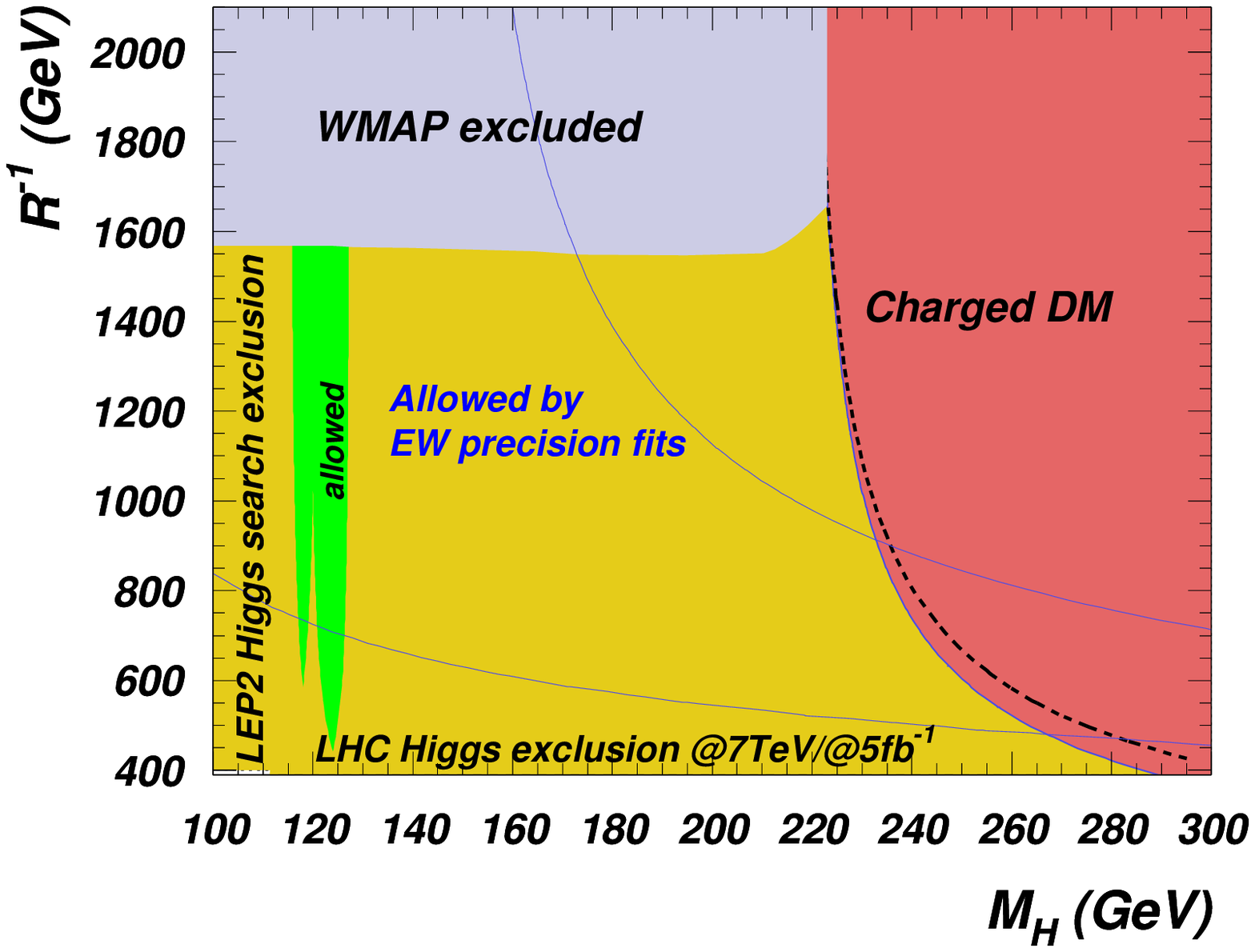} 
	\end{center}
    \caption{Left: exclusion of mUED $(m_h,R^{-1})$ parameter space at 95\% CL
    from Higgs boson search using combined ATLAS and CMS limits in
    $\gamma\gamma$, $W^+ W^-$ and $ZZ$ channels, based on the 7~TeV data. The allowed
    region is in light green and the excluded region is in red. Dark
    green shows the additional allowed region when loop corrected KK masses are
    used	instead of tree-level masses. Contours of constant
    $\mu^{95\%}$ are shown in steps of 0.05.
    Right: Combination of limits on the mUED parameter space from: the Higgs
    constraints considered in this paper; EW precision tests (95\% CL); and DM
    relic density limits for $\Lambda=40 R^{-1}$ (solid line) and $\Lambda=20
    R^{-1}$ (dashed line) cases.}
    \label{fig:mued-limits}
\end{figure}

Using our model's predictions of Higgs production enhancement for different
values of $m_{h}$ and $R^{-1}$ together with experimental limits on Higgs boson
production, we can exclude regions of the $(m_h,R^{-1})$ plane where
$\mu^{95\%} < 1$. Initially, we statistically combined the CMS data from Fig.~6
(top) of \cite{Chatrchyan201226} and the ATLAS data from Fig.~3 of
\cite{ATLAS-CONF-2012-019} in each of the $\gamma\gamma$, $W^+W^-$ and $ZZ$
channels. Note that these data are from the old 7~TeV dataset, before the
discovery of a Higgs-like particle at 125~GeV in July 2012. When we started writing this paper, this
was the state of the art. We update the analysis using the newest 8~TeV data
later in this section.

The resulting limits on mUED from the 7~TeV dataset are shown in our
Fig.~\ref{fig:mued-limits} (left),
where the  green contour separating the green and red shaded regions
corresponds to $\mu^{95\%}=1$ level.
The  other contours of constant $\mu^{95\%}$ are shown  in steps of 0.05 for increasing value of $\mu^{95\%}$ towards the green region and its decreasing value in the opposite direction.
The red-shaded region of the parameter space is excluded at 95\% confidence level.
These constraints are combined with other
constraints from DM relic density~\cite{Belanger:2010xy} as well as EW
precision tests~\cite{Gogoladze:2006br} in Fig.~\ref{fig:mued-limits} (right).

We can see that Higgs searches powerfully constrain mUED, in which Higgs
production is enhanced.  Compared to previous studies \cite{Nishiwaki:2011gk}
we  have included mass corrections for the particles in the loops, providing
more realistic predictions of mUED cross sections, and have accurately combined
non-universal enhancement for $\gamma\gamma$ and $W^+ W^-$/$ZZ$ signatures.

This new approach allows us to find accurate limits on the mUED $(m_h,R^{-1})$
parameter space. After combination of ATLAS and CMS limits for each individual
channel ($\gamma\gamma$,  $W^{+}W^{-}$ and $ZZ$) in gluon-gluon fusion, we find
that $R^{-1}<500$~GeV is   excluded at 95\%CL.  For $500 \text{ GeV}< R^{-1}<600
\text{ GeV}$ only a very narrow ($\pm 1-3$~GeV) mass window around $m_h=125$~GeV
is left. This is the region where  the excess of the events in the Higgs search
channels is reported by the ATLAS and CMS collaborations and where the
exclusion limit is weaker. For even larger values of $R^{-1}$ another narrow
mass range around $m_h=118$~GeV is allowed. 

\begin{figure} 
    \begin{center}
            \includegraphics[width=0.6\textwidth]{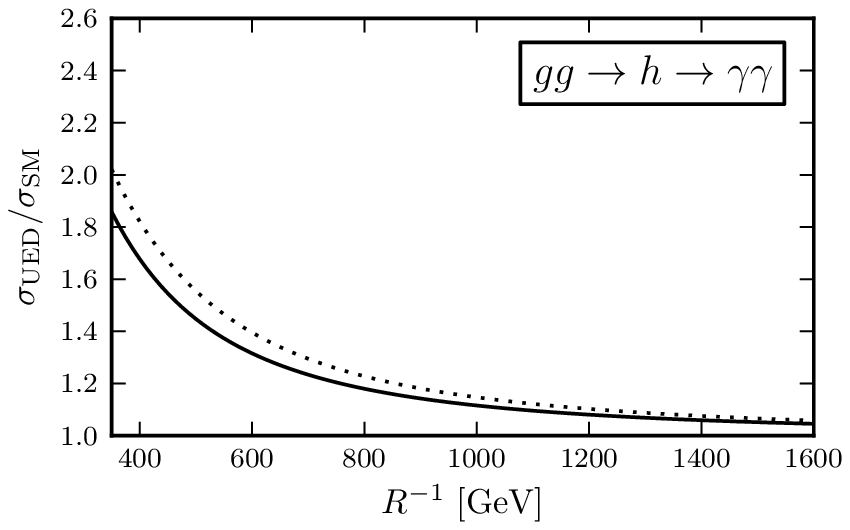}\\
            \includegraphics[width=0.6\textwidth]{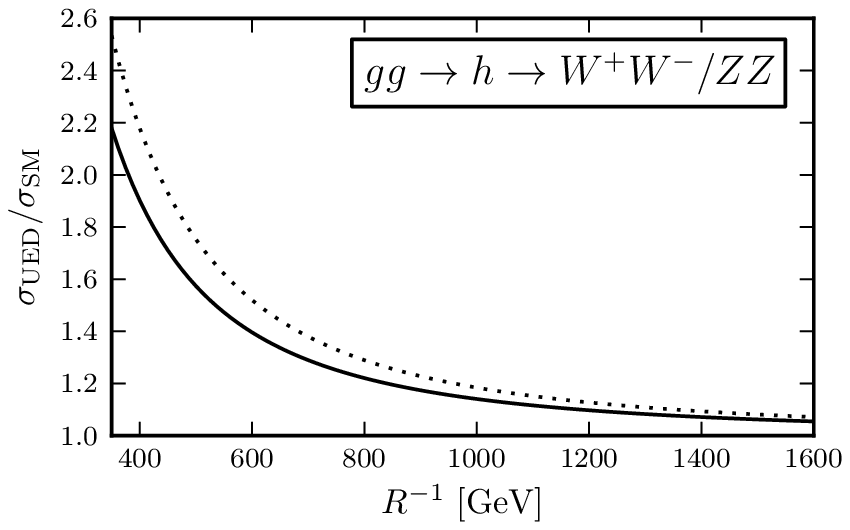}\\
            \includegraphics[width=0.6\textwidth]{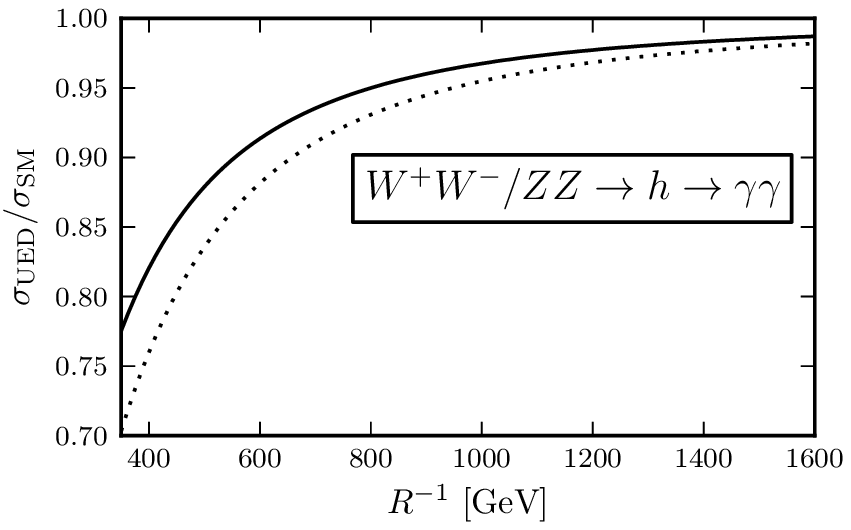}
    \end{center}
    \caption{
     The variation with respect to $R^{-1}$ of the mUED cross-sections for 
     $gg \to h \to \gamma\gamma$ (top), $gg \to h \to W^+W^-/ZZ$ (middle) 
     and $W^+W^- /ZZ \to h \to \gamma\gamma$ (bottom) 
     channels relative to the SM for $m_h = 125$~GeV. Solid lines show 
    results when using loop-corrected masses in the loops, while dashed lines 
    correspond to tree-level masses.}
    \label{fig:fixedmh}
\end{figure}

For a Higgs mass $m_h=125$~GeV, we display in Fig.~\ref{fig:fixedmh} the
variation of the enhancement factor  in the  $gg\rightarrow h\rightarrow
\gamma\gamma$ (top) and $gg\rightarrow h\rightarrow W^{+}W^{-}/ZZ$ (middle)
channels as a function of $R^{-1}$ together with the suppression factor  in the
$W^+ W^- /ZZ \rightarrow h\rightarrow \gamma\gamma$ (bottom). The latter is
relevant for the Higgs search in the $pp\rightarrow jj\gamma\gamma$.   These
plots can be used to ascertain how
a measurement of each channel's cross-section can be used to constrain the scale $R^{-1}$. For
example, an enhancement in both  the $gg\rightarrow h\rightarrow \gamma\gamma$
and the $gg\rightarrow h\rightarrow W^{+}W^{-}$ channel would favour the mUED
model around the TeV scale while a large enhancement in  $pp\rightarrow
jj\gamma\gamma$ would disfavour the model.

Since we started to write this paper, new limits (calculated from the first tranche of
8~TeV data) have been released by CMS~\cite{Chatrchyan:2012gu} and
ATLAS~\cite{Aad:2012gk}. The data are strong enough for each experiment to claim
discovery of a Higgs-like particle with a mass of around 125~GeV, confirming
the hints evident in earlier analyses. The new ATLAS limits are shown for all
channels in Fig.~16a of the supplementary figures associated with
\cite{Aad:2012gk}.\footnote{These can be found at
    \url{https://atlas.web.cern.ch/Atlas/GROUPS/PHYSICS/PAPERS/HIGG-2012-27/}}
CMS make their latest limits for $\gamma\gamma$ available in Fig.~4a of
\cite{CMS-PAS-HIG-12-015} and their limits for $WW$ in Fig.~4 (right) of
\cite{Chatrchyan:2012gu}. The CMS limits for the $h \to ZZ \to 4\ell$ channel can be
found in the supplementary figures for
\cite{CMS-PAS-HIG-12-016}.\footnote{\url{https://twiki.cern.ch/twiki/bin/view/CMSPublic/Hig12016TWiki}}

\begin{figure}
    \begin{center}
        \includegraphics{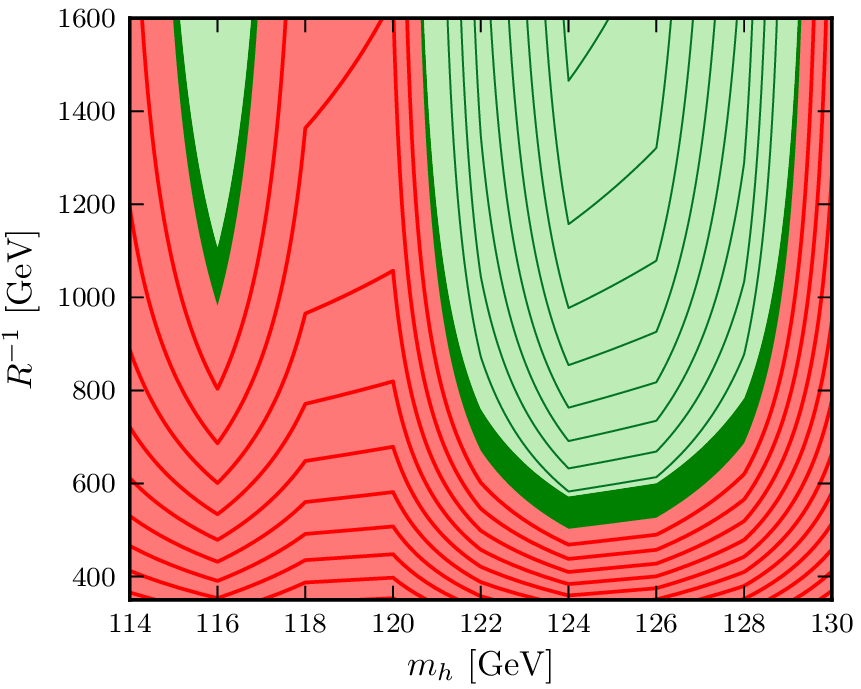}
        \includegraphics{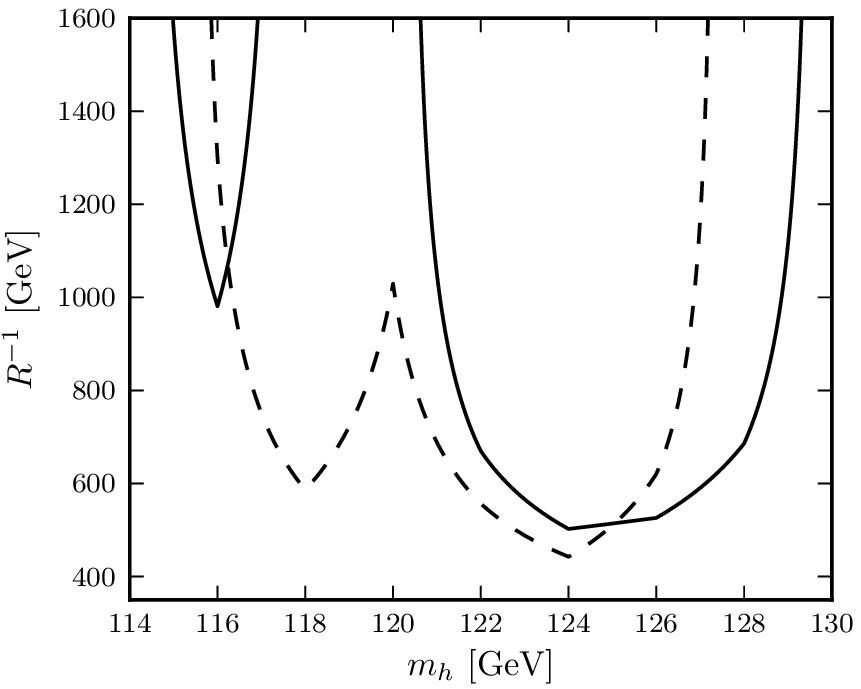}
    \end{center}
    \caption{Left: limits on mUED parameter space from newest 7 TeV and 8 TeV ATLAS
    and CMS Higgs search data using the same conventions as in 
    Fig.~\ref{fig:mued-limits} (left).
    Right: comparison of allowed regions for the combined 7~TeV and 8~TeV LHC
    data (solid) and 7~TeV data (dashed) using loop masses.}
    \label{fig:limit_comparison}
\end{figure}

We have calculated the constraints on the mUED parameter space in light of
these new experimental data and the result is shown in
Fig.~\ref{fig:limit_comparison} (left). We also show a comparison of the
allowed regions for the old and new data in Fig.~\ref{fig:limit_comparison}
(right).

We should also comment on the expected sources of uncertainty in our approach.
Since our study is  based on the {\it ratio} of mUED and SM cross-sections our 
results are insensitive to PDF uncertainties which simply cancel in this ratio.
The other potential sources of uncertainty are the higher order corrections to the amplitudes we calculate.
Fortunately,  higher order corrections has been evaluated 
for $h\to gg$ process to four loops in [44].
Using the results from that paper one can estimate that the biggest uncertainty in our results
from higher order corrections 
comes from the second loop term, containing an additional $\log(m_q^{(n)}/m_h)$ dependence due to mUED. It turns out numerically that
this effect  is about $1\%\times (\sigma_{mUED}/\sigma_{SM}) < 1\%$
and is thus negligible. 
Therefore the biggest source of uncertainty is actually related to the choice of using
  loop-corrected versus tree-level masses one in our loop calculations.
As we argue above, we choose the loop-corrected mass for our evaluations, but in order to be on the conservative side we  consider the impact of choosing the tree-level mass instead. We use this difference to estimate the uncertainty in our limits. With tree-level masses, our limits 
presented  in both
Fig.~4 and Fig.~6 are  shifted by about 50~GeV. In fact the limits actually improve when using the tree-level masses.

The allowed region shrinks overall with the extra data, but the high and low
$m_h$ limits on $R^{-1}$ \emph{relax} down to about 550 GeV. This is actually to be expected: in the 2011 data,
the $W^+W^-$ channel surprisingly showed no excess of events around 125 GeV
even though such an excess was observed in the other channels, including $ZZ$.
In the new data, there is an excess in $W^+W^-$, bringing this channel in line
with the others and thus \emph{weakening} the limit on the mUED parameter space
slightly at the edges of the allowed region where the diphoton channel is less
restrictive. However, the improvement in limiting power of the diphoton channel
causes the region $117\text{ GeV}\lesssim m_h \lesssim 121\text{ GeV}$ to become
forbidden.

With the new data then, all values of $R^{-1} < 550 \text{ GeV}$ are
forbidden, leaving a small region of allowed parameter space 2--8 GeV wide
around $m_h = 125 \text{ GeV}$ and another allowed island up to 2~GeV wide
around 116~GeV for $R^{-1} > 1000\text{ GeV}$.

%% file: 06-conclude-mued-higgs.tex
%!TEX root = /Users/matt/Documents/Extra dimensions/Higgs/dropbox_contents/higgs_paper/mued-higgs.tex

\section{Conclusions}
\label{sec:conclusions}

LHC searches for the SM Higgs provide a powerful limit on  mUED model where the Higgs production is enhanced.
We have evaluated  all one-loop diagrams for Higgs production 
$gg\to h$ and decay $h\to \gamma\gamma$ within the mUED model and have 
independently confirmed previous results~\cite{Petriello:2002uu}.
Based on these results we have derived enhancement  factors for Higgs boson production and decay in the  mUED parameter space. 
%and have used these factors  to find the limit on the mUED $(m_h,R^{-1})$ parameter space.
Then, using these factors we have derived the first limits on 
the  mUED parameter space
which combine both limits from ATLAS and CMS collaborations
for  7~TeV and 8~TeV LHC data
and take into account statistical combination of several Higgs boson search channels properly.
As for other extensions of the SM, the correct  statistical combination of several Higgs boson search channels is important for mUED  since these channels are not universally enhanced: the 
 $gg\to h \to \gamma\gamma$ process is not enhanced as strongly as  the
 $gg\to h \to WW^*$ or $gg\to h \to ZZ^*$  processes due to the fact that the decay
 $h \to \gamma\gamma$ is actually suppressed as compared to the Standard Model.
Overall enhancement for $gg\to h \to \gamma\gamma$ nevertheless takes place because the  enhancement of $gg\to h$ overcomes the suppression in the $h \to \gamma\gamma$ decay.

In contrast to previous studies \cite{Nishiwaki:2011gk} we have included mass
corrections for the KK particles in the loop and found that the effect of KK
particles is slightly reduced as compared to the calculation using tree-level
masses. The comparison between the computations with tree-level and radiatively
corrected masses provides information about the theoretical uncertainties in
the enhancement of the Higgs boson production and decay within the mUED model.
Also, we think that including these mass corrections gives more precise result
and allows one to take into account some part of the higher order corrections.
This is since one-loop corrected masses give a better approximation to pole
masses and since the coupling constants that couple the gluon (or photon) to
the KK quarks are protected by gauge invariance from receiving radiative
corrections.

% $gt_{1,2}^{(n)}t_{1,2}^{(n)}$ and $\gamma t_{1,2}^{(n)}t_{1,2}^{(n)}$ coupling constants do not receive radiative corrections due to gauge invariance.

As a result we have found an accurate limit on mUED in the $(m_h,R^{-1})$
parameter space. After combination of ATLAS and CMS limits for each individual
channel ($\gamma\gamma$,  $WW^*$ and $ZZ^*$)  for the latest 7~TeV and 8~TeV data, we found that $R^{-1}<550$~GeV is
excluded at 95\%CL, while for larger  $R^{-1}$ only a very narrow ($\pm
1-4$~GeV) mass window around $m_h=125$~GeV (the mass of the recently observed Higgs-like particle), and another smaller window around 118~GeV (for $R^{-1} > 1000\text{ GeV}$) remain allowed. 

As new 8~TeV data becomes available, the results from the different Higgs search channels can be used to fit the mUED parameter space.  Signals compatible with the SM would eventually push the values of $R^{-1}$ above the TeV scale while deviations from the SM could either be compatible with a lower scale or even exclude  mUED completely depending on the channels involved. Indeed  mUED predicts 
an enhancement for all  channels for $gg\to h$ production and decay. On the other hand, the vector boson fusion process $WW/ZZ \to h \to \gamma\gamma$
is generically suppressed in mUED while $WW/ZZ \to h \to WW^*/ZZ^*$ is standard. A confirmation of the larger excess in the vector boson fusion mode over the gluon fusion mode for
the two-photon channel that is currently observed would disfavour mUED. On the
other hand predictions that come closer to the SM ones would lead to an
increase in the mUED scale.

With detailed information on individual Higgs  boson  production and decay processes  provided by CMS and ATLAS experiments, one can  understand  much better  the  nature  of the Higgs  boson and interpret it within  mUED or other BSM theories.

%% file: appendix-mued-higgs.tex
%!TEX root = /Users/matt/Documents/Extra dimensions/Higgs/dropbox_contents/higgs_paper/mued-higgs.tex
\appendix

\section{Feynman rules}
\label{app:feynmanrules}

Below is a table of the Feynman rules for the propagators and vertices needed
to evaluate the diagrams contributing to the $gg \to h$ and $h \to
\gamma\gamma$ amplitudes. The vertex rules are given in terms
of a general coefficient; underneath this, the value of the coefficient is
written for the SM case and for the $n$th KK level. We use a $(+---)$
signature and the following momentum conventions: fermion momentum flows in
the same direction as fermion number and external momentum flows inwards. This
convention is shown graphically in Fig.~\ref{fig:feyndia}

% \newcolumntype{M}{m{0.24\textwidth}}
\begin{longtable}{m{0.24\textwidth}m{0.24\textwidth}m{0.24\textwidth}m{0.24\textwidth}}
  \hline\hline
  \begin{equation*}
    \begin{array}{c}\includegraphics[scale=0.75]{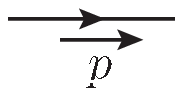}\end{array}
    = \frac{\ii (\slashed{p}+m)}{p^2-m^2+\ii\epsilon}
  \end{equation*} &
  \begin{equation*}
    \begin{array}{c}\includegraphics[scale=0.75]{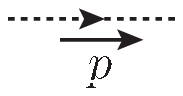}\end{array}
    = \frac{\ii}{p^2-m^2+\ii\epsilon}
  \end{equation*} & 
  \begin{equation*}
    \begin{array}{c}\includegraphics[scale=0.75]{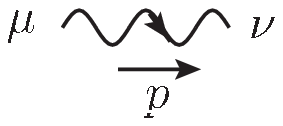}\end{array}
    = \frac{-\ii g_{\mu\nu}}{p^2-m^2+\ii\epsilon} \;\text{(Feynman gauge)}
  \end{equation*} \\
  \hline
  \begin{center}\includegraphics[scale=0.75]{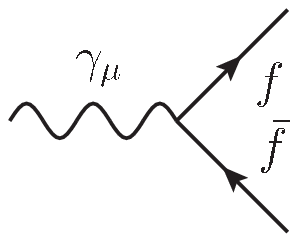}\end{center} &
  \begin{gather*}
    \ii G_{ff}\gamma_\mu \\
    G_{ff}^{\text{SM}} = -e Q_f \\
    G_{ff}^{(n)} = -e Q_f
  \end{gather*} &
  \begin{center}\includegraphics[scale=0.75]{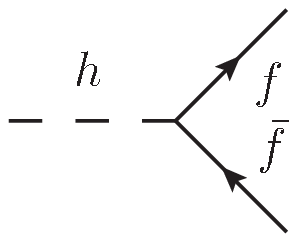}\end{center} &
  \begin{gather*}
    \ii \lambda_{ff} \\
    \lambda_{ff}^{\text{SM}} = -\frac{g m_f}{2m_W} \\
    \lambda_{ff}^{(n)} = - \frac{g m_f}{2m_W} \sin 2a_f^{(n)}
  \end{gather*} \\
  \hline
  \begin{center}\includegraphics[scale=0.75]{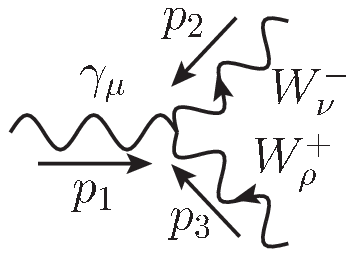}\end{center} &
    \multicolumn{3}{m{0.4\textwidth}}{
      \begin{gather*}
        \ii G_{WW}( p_{3\mu}g_{\nu\rho}
        -p_{3\nu}g_{\mu\rho}-p_{1\rho}g_{\mu\nu}
        +p_{1\nu} g_{\mu\rho}+p_{2\rho}g_{\mu\nu}-p_{2\mu}g_{\nu\rho} ) \\
        G_{WW}^{\text{SM}} =-e \\
        G_{WW}^{(n)} = -e
      \end{gather*}
    } \\
  \multicolumn{2}{m{0.42\textwidth}}{
    \begin{equation*}
      \begin{array}{c}\includegraphics[scale=0.75]{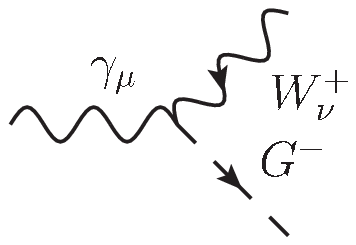}\end{array} =
      -\left(\begin{array}{c}\includegraphics[scale=0.75]{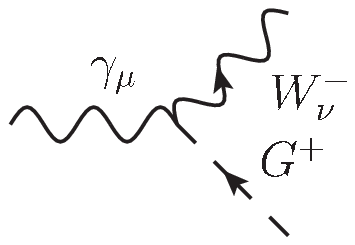}\end{array}\right)
    \end{equation*}
  } &
  \begin{gather*}
    G_{WG} g_{\mu\nu} \\ 
    G_{WG}^{\text{SM}} = e m_{W} \\
    G_{WG}^{(n)} = e m_{W,n} \\
  \end{gather*} & \\
  \begin{center}\includegraphics[scale=0.75]{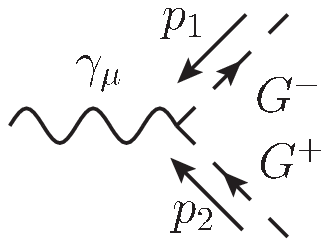}\end{center} &
    \begin{gather*}
      \ii G_{GG}  (p_{2}-p_{1})_{\mu} \\
      G_{GG}^{\text{SM}} = e \\
      G_{GG}^{(n)} = e
    \end{gather*} &
  \begin{center}\includegraphics[scale=0.75]{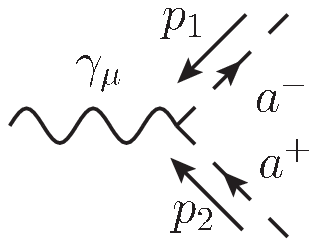}\end{center} &
    \begin{gather*}
      \ii G_{aa} (p_{2}-p_{1})_{\mu} \\
      G_{aa}^{\text{SM}} = 0 \\
      G_{aa}^{(n)} = e
    \end{gather*} \\
  \multicolumn{2}{m{0.48\textwidth}}{
    \begin{equation*}
      \begin{array}{c}\includegraphics[scale=0.75]{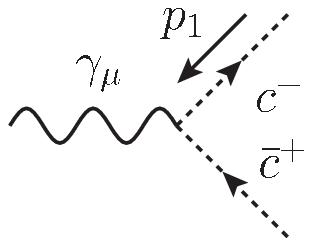}\end{array}=
      -\left(\begin{array}{c}\includegraphics[scale=0.75]{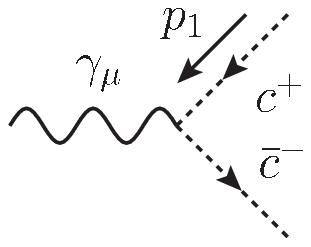}\end{array}\right)
    \end{equation*}
  } &
    \begin{gather*}
      \ii G_{\bar{c}c} p_{1\mu} \\
      G_{\bar{c}c}^{\text{SM}} = -e \\
      G_{\bar{c}c}^{(n)} = -e
    \end{gather*} &  \\
  \hline
  \begin{center}\includegraphics[scale=0.75]{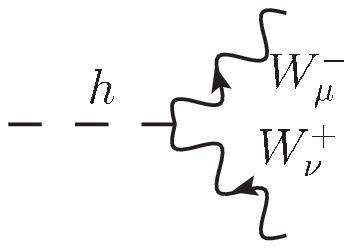}\end{center} & 
    \begin{gather*}
      \ii\lambda_{WW}g_{\mu\nu} \\
      \lambda_{WW}^\text{SM} = g m_W \\
      \lambda_{WW}^{(n)} = g m_W
    \end{gather*} &
  \begin{center}\includegraphics[scale=0.75]{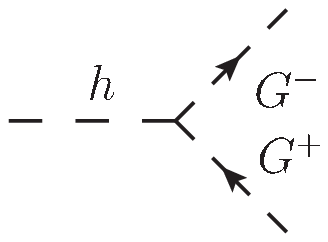}\end{center} &
    \begin{gather*}
      \ii \lambda_{GG} \\
      \lambda_{GG}^\text{SM} = -gm_h^2/(2m_W) \\
      \lambda_{GG}^{(n)} = -\frac{g m_h^2}{2m_W}
        \left(\frac{m_W}{m_{W,n}}\right)^2
    \end{gather*} \\
  \multicolumn{2}{m{0.48\textwidth}}{
    \begin{equation*}
      \begin{array}{c}\includegraphics[scale=0.75]{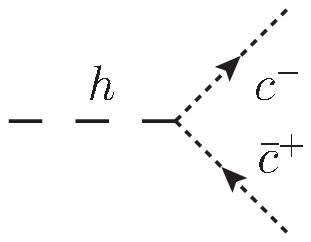}\end{array} =
      \begin{array}{c}\includegraphics[scale=0.75]{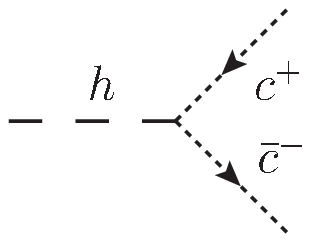}\end{array}
    \end{equation*}
  }&
  \begin{gather*}
    \ii \lambda_{\bar{c}c} \\
    \lambda_{\bar{c}c}^{\text{SM}} = -g m_W /2 \\
    \lambda_{\bar{c}c}^{(n)} = -g m_W / 2\end{gather*} & \\
  \multicolumn{2}{m{0.48\textwidth}}{
    \begin{equation*}
      \begin{array}{c}\includegraphics[scale=0.75]{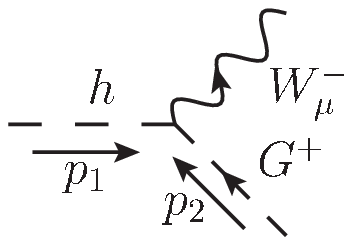}\end{array} =
      \begin{array}{c}\includegraphics[scale=0.75]{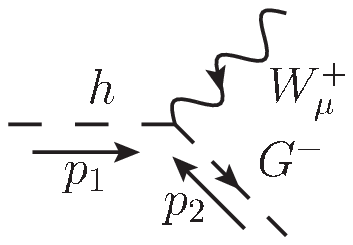}\end{array}
    \end{equation*}
  } &  
  \begin{gather*}
    -\lambda_{WG} (p_{2}-p_{1})_{\mu} \\
    \lambda_{WG}^{\text{SM}}=-g/2 \\
    \lambda_{WG}^{(n)}=-(g/2)(m_W/m_{W,n})
  \end{gather*} &  \\  
  \begin{center}\includegraphics[scale=0.75]{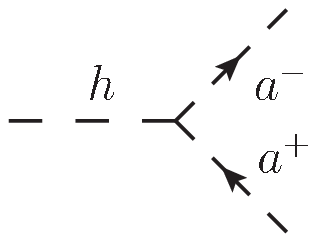}\end{center} &
    \multicolumn{3}{m{0.4\textwidth}}{
      \begin{gather*}
        \ii\lambda_{aa} \\
        \lambda_{aa}^{\text{SM}}=0 \\
        \lambda_{aa}^{(n)}=-\frac{g}{2m_W}\left[2\left(\frac{m_{a,n}}{m_{W,n}}
          \right)^2 m_W^2 + m_h^2\left(1-\frac{m_W^2}{m_{W,n}^2}\right)\right]
      \end{gather*}
    }\\
    \hline
    \begin{center}\includegraphics[scale=0.75]{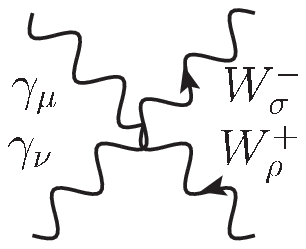}\end{center} & 
    \multicolumn{3}{m{0.4\textwidth}}{
      \begin{gather*}
        \ii H_{WW} (2 g_{\mu\nu} g_{\rho\sigma}-g_{\mu\sigma} g_{\nu\rho} - 
        g_{\mu\rho} g_{\nu\sigma}) \\
        H_{WW}^{\text{SM}} = -e^2 \\
        H_{WW}^{(n)} = -e^2
      \end{gather*}
    } \\
    \begin{center}\includegraphics[scale=0.75]{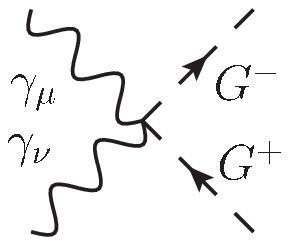}\end{center} & 
    \begin{gather*}
      \ii H_{GG} g_{\mu\nu} \\
      H_{GG}^{\text{SM}} = 2e^2 \\
      H_{GG}^{(n)} = 2e^2
    \end{gather*} &
    \begin{center}\includegraphics[scale=0.75]{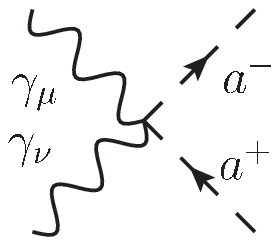}\end{center} & 
    \begin{gather*}
      \ii H_{aa} g_{\mu\nu} \\
      H_{aa}^{\text{SM}} = 0\\
      H_{aa}^{(n)} = 2e^2
    \end{gather*} \\
    \hline
    \multicolumn{2}{m{0.48\textwidth}}{
      \begin{equation*}
        \begin{array}{c}\includegraphics[scale=0.75]{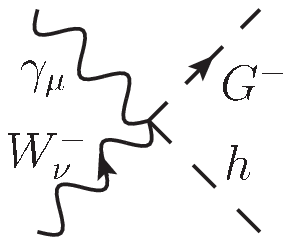}\end{array}
        \,=\,
        -\;\begin{array}{c}\includegraphics[scale=0.75]{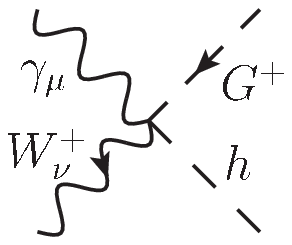}\end{array}
      \end{equation*}
    } &
    \begin{gather*}
      I_{WG} g_{\mu\nu} \\
      I_{WG}^{\text{SM}} = -\frac{e g}{2}\\
      I_{WG}^{(n)} = -\frac{e g}{2} \frac{m_W}{m_{W,n}}
    \end{gather*} & \\
    \hline\hline \\
\end{longtable}

\section{$gg\to h$ amplitude}
\label{app:ggh}

Here we calculate the generic 1-loop amplitude
for two gluons to produce a Higgs boson via a quark loop. We leave the
couplings and quark masses general for now and will specialise to the SM and
mUED case below.
\begin{equation*}
  \ii(\mathcal{A}_{ggh,q})^{ab}_{\mu\nu} = 
2\times\;
\begin{array}{c}
  \includegraphics[scale=0.5]{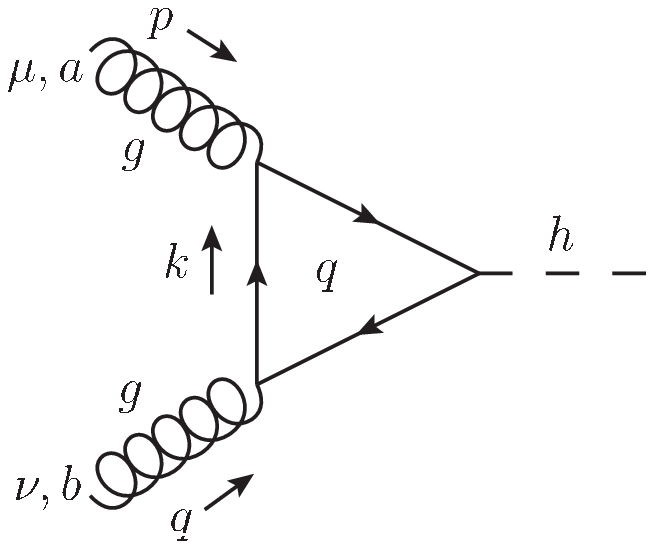}
\end{array}
\end{equation*}
The factor of two is to count the diagram formed by swapping the external
gluons. To form the amplitude above diagram should be contracted with the gluon
polarisation vectors which carry Lorentz and group indices. The labeled arrows
denote momentum flow; the other labels designate the particle names.

The KK and SM quarks couple to gluons identically as $\ii g_{s} t_{a}$, where
$g_s$ is the strong coupling constant and $t^a$ is an SU(3) generator. We call
the Yukawa coupling $\ii\lambda_{q}$.

Performing the loop momentum integral in $D=4-\epsilon$ dimensions to regulate
the divergence (and introducing the renormalisation scale $\mu$ to
compensate), the amplitude without polarisation vectors is
\begin{align*}
\ii(\mathcal{A}_{ggh,q})_{\mu\nu}^{ab} &= -2(\ii
g_{s})^{2}(\ii\lambda_{q})\text{Tr}(t^{a}t^{b}) \mu^{4-D}\\
&\quad \times \int \frac{\d^{D}k}{(2\pi)^{D}}
\text{tr}\bigg\{\frac{\ii(\slashed{k}+m)}{k^{2}-m^{2} + \ii\epsilon}
\gamma_{\mu} \frac{\ii (\slashed{k} + \slashed{p} +
m)}{(k+p)^{2}-m^{2} + \ii\epsilon} \frac{\ii(\slashed{k} -
\slashed{q}+m)}{(k-q)^{2}-m^{2} + \ii\epsilon}\gamma_{\nu}\bigg\}
\end{align*}
with the overall minus sign due to the fermion loop. The propagator
conventions used here are given in appendix~\ref{app:feynmanrules} with the
mass set to a general quark mass $m$ (we reserve the symbol $m_q$ for the SM
mass of quark $q$). The trace Tr is over the SU(3) generators and tr is over
the product of Dirac matrices.

The rest of the calculation (and all following calculations) assumes that the
gluons and Higgs boson are physical, so $p^2 = q^2 = 0$,
$p^\mu\epsilon_\mu(p) = q^\nu \epsilon_\nu(q) = 0$ and $(p+q)^2 = m_h^2$. The
approximation that the Higgs is real is justified if the ``narrow width
approximation'' is valid (see Sec.~\ref{sec:calcs} for details).

The numerator of the Dirac trace (rejecting off-shell terms as discussed above) is
\[
\ii^3\times 4 m[g_{\mu\nu}(m^{2}-k^{2}-m_{h}^{2}/2)+4k_\mu k_\nu+p_{\nu}q_{\mu}] = \ii^3\times 4 m\{
[(m^{2}-m_{H}^{2}/2)g_{\mu\nu}+ p_{\nu}q_{\mu}]-g_{\mu\nu}k^{2} + 4k_{\mu}k_{\nu}\}.
\]
In terms of PV functions the amplitude becomes
\[
\ii(\mathcal{A}_{ggh,q})_{\mu\nu}^{ab}= -2(\ii g_{s})^{2}(\ii\lambda_{q})\text{Tr}(t^{a}t^{b})\ii^{3} \frac{\ii\pi^{2}}{(2\pi)^{4}}4m\left\{
[(m^{2}-m_{h}^{2}/2)g_{\mu\nu}+p_{\nu}q_{\mu}]C_{0} + g_{\mu\nu}C^{\rho}_{\rho}+4C_{\mu\nu}
\right\}.
\]
Performing Passarino-Veltman reduction, and carefully taking the limit $D\to 4$, we find that
\[
\ii(\mathcal{A}_{ggh,q})_{\mu\nu}^{ab} = \frac{\ii}{2\pi^{2}} \lambda_{q} g_{s}^{2} \text{Tr}(t^{a}t^{b}) m \left(\frac{g_{\mu\nu}m_{h}^{2}}{2} - p_{\nu}q_{\mu}\right)\left[\frac{2}{m_{h}^{2}}-\left(1-\frac{4m^{2}}{m_{h}^{2}}\right)C_{0}\right].
\]
For SU(3) generators, $\mathrm{Tr}(t^{a}t^{b})=\frac{1}{2}\delta^{ab}$ so the
quark $q$'s total contribution to the amplitude is
\[
\ii(\mathcal{A}_{ggh,q})_{\mu\nu}^{ab} = \frac{\ii\alpha_{s}}{\pi} \delta^{ab}
\left(\frac{g_{\mu\nu}m_{h}^{2}}{2} - p_{\nu}q_{\mu}\right) \lambda_{q} m
\left[\frac{2}{m_{h}^{2}}- \left(1 - \frac{4m^{2}}{m_{h}^{2}}\right)
C_{0}(m,m_{h}) \right],
\]
where $\alpha_s = g_s^2/4\pi$.

It is useful to factor out the Lorentz and colour dependence by defining the
``reduced amplitude'' $\tilde{\mathcal{A}}$ for a particular process in terms of the
the full (\emph{sans} polarisation vectors) amplitude
$\mathcal{A}_{\mu\nu}^{ab}$:
\[
\mathcal{A}_{\mu\nu}^{ab} =
\tilde{\mathcal{A}}\times\delta^{ab}\left(\frac{g_{\mu\nu}m_{h}^{2}}{2} -
p_{\nu}q_{\mu}\right),
\]
so in this case
\[
\tilde{\mathcal{A}}_{ggh,q} = \frac{\alpha_s}{\pi}\lambda_q m \left[\frac{2}{m_h^2} -
\left(1-\frac{4m^2}{m_h^2}\right)C_0(m,m_h)\right],
\]
which can be written in terms of the function defined in \eqref{eq:fF} as
\begin{equation}
\tilde{\mathcal{A}}_{ggh,q} = \frac{\alpha_s}{4\pi}\lambda_q \frac{1}{m} f_F(m).
\label{eq:agghq}
\end{equation}

\subsection{The SM and mUED cases}
\subsubsection{Standard Model}
Equation~\eqref{eq:agghq} is in terms of the mass $m$ and Yukawa coupling
$\lambda_q$ of a general quark $q$. For the SM quarks, let $m = m_{q}$ with $q
\in \{u,d,s,c,b,t\}$. The SM Yukawa coupling in terms of the Higgs vacuum
expectation value (VEV) $v$ is $\lambda_{q}^{\text{SM}} = -m_{q}/v$, so
\[
\tilde{\mathcal{A}}_{ggh}^{\text{SM}} = -\frac{\alpha_{s}}{4\pi v}F_{ggh}^\text{SM},
\]
where
\begin{equation}
  F_{ggh}^{\text{SM}} = \sum_{q}f_{\text{F}}(m_{q}),
  \label{eq:fgghsm}
\end{equation}
which is the expression shown in \eqref{eq:gghmaster} and the following paragraph in Sec.~\ref{sec:calcs}.

\subsubsection{Including KK modes}
At each KK level $n$, there are two types of quarks $q_1^{(n)}$ and
$q_2^{(n)}$ for each SM quark $q$. At tree level, these quarks' masses would
both be $\sqrt{m_q^2 + n^2/R^2}$, but if one-loop mass corrections are included then they
split. However, Yukawa couplings to the Higgs are shifted equally under mass
corrections so $\lambda_{q}^{(n)} = -m_{q} \sin (2a_{q}^{(n)})/v$ for both
$q_1^{(n)}$ and $q_2^{(n)}$. Here $a_q^{(n)}$ is the mixing angle between
quark flavour eigenstates $(q_L^{(n)},q_R^{(n)})$ and mass eigenstates
$(q_1^{(n)},q_2^{(n)})$; this is explained further in \cite{belyaev:2012a}.

The contribution to $\tilde{\mathcal{A}}_{ggh}$ from the KK level $n$ quarks
is then $\tilde{\mathcal{A}}_{ggh}^{(n)} = -\frac{\alpha_s}{4\pi
v}F_{ggh}^{(n)}$, where
\[
F_{ggh}^{(n)} = \sum_q \sin (2a_q^{(n)}) \left(\frac{m_q}{m_{q,1}^{(n)}}
f_F(m_{q,1}^{(n)},m_h) + \frac{m_q}{m_{q,2}^{(n)}}f_F(m_{q,2}^{(n)},m_h)
\right).
\]
The full expression for $F_{ggh}$ (and hence $\tilde{\mathcal{A}}_{ggh}$), as given in \eqref{eq:fgghfull}, is obtained by summing over the KK number $n$ and adding the SM contribution \eqref{eq:fgghsm}.

\section{$h\to\gamma\gamma$ amplitude}
\label{app:haa}

The full $h\to\gamma\gamma$ amplitude receives contributions from fermions
(quarks and leptons), $W$ bosons and charged scalars $a^\pm$ (which appear at
KK number 1 and above). We use the subscript $f$, $W$ and $a$ to distinguish
these contributions.

\subsection{Fermion contribution}
For each fermion there are two contributing diagrams (equal to each other and related by the swapping external photons).
\[
    \ii(\mathcal{A}_f)_{\mu\nu} = 2\times\; 
    \begin{array}{c}
        \includegraphics[scale=0.75]{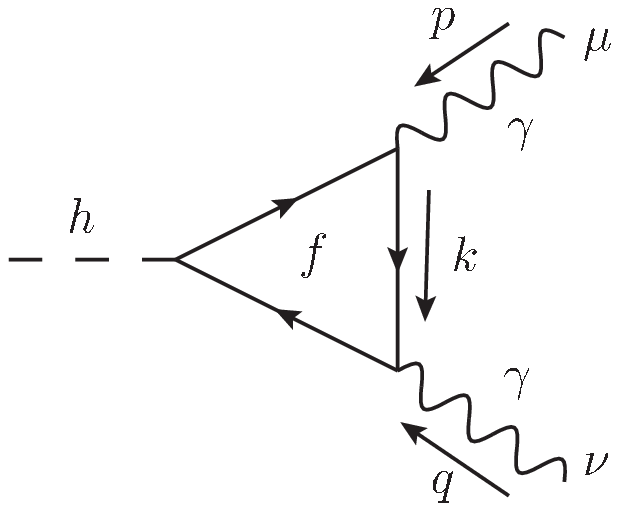}
    \end{array}
\]
Leaving the couplings general and using the Feynman rules in Appendix~\ref{app:feynmanrules} we find that
\begin{align*}
\ii (\mathcal{A}_f)_{\mu\nu} &= -2(+\ii
G_{ff})^{2} (+\ii\lambda_{ff}) (+\ii)^{3} \int_k \frac{1}{\mathcal{D}}\text{tr} [(\slashed{k}+m) \gamma_{\mu} (\slashed{k} +
\slashed{p} + m)(\slashed{k} - \slashed{q} + m)
\gamma_{\nu}] \\
&= \frac{2\ii G_{ff}^{2}\lambda_{ff}}{16\pi^{2}}\left[
4\frac{m}{m_{h}^{2}}(4m^{2}-m_{h}^{2})C_{0}+ \frac{8
m}{m_{h}^{2}}\right]\left(\frac{m_{h}^{2}g_{\mu\nu}}{2} -
p_{\nu}q_{\mu}\right).
\end{align*}
where we have used the shorthand
\[
\int_k \equiv \int \frac{\d^D k}{(2\pi)^D}\mu^{4-D} 
\]
for the dimensionally-regularised momentum integral and where we have written the denominator, common to all triangle diagrams considered in this paper, as
\[
\mathcal{D} = [k^2-m^2+\ii\epsilon] [(k+p)^2-m^2+\ii\epsilon]
[(k-q)^2-m^2+\ii\epsilon].
\]

Factoring out the Lorentz part yields as in the $ggh$ case leaves
\[
\tilde{\mathcal{A}}_f = \frac{G_{ff}^2 \lambda_{ff}}{8\pi^2}\frac{1}{m} f_F(m),
\]
with $f_F(m)$ defined as in \eqref{eq:fF}.

Specialising to the SM using the rules in Appendix~\ref{app:feynmanrules} gives
\[
\tilde{\mathcal{A}}_f^{\text{SM}} = -\frac{Q_f^2 e^2}{8\pi^2 v}f_F(m_f),
\]
where $v=2m_W/g$ is the Higgs VEV and $Q_f e$ is the charge of the fermion.

The contribution from an $n$th level KK fermion is
\[
\tilde{\mathcal{A}}_f^{(n)} = -\frac{Q_f^2 e^2}{8\pi^2 v} \sin 2a_f^{(n)}
f_F(m_f^{(n)}),
\]
where $a_f^{(n)}$ is the mixing angle for converting from the flavour to the mass eigenbasis of the KK fermion.

\subsection{Gauge boson contribution}
There is an additional (in fact dominant) contribution to the
$h\to\gamma\gamma$ amplitude from SM and KK $W$ bosons and their associated
Goldstone bosons and Faddeev-Popov ghosts. We chose to perform the calculation
in the 't Hooft-Feynman gauge (the $R_\xi$ gauge with $\xi=1$). The relevant
diagrams, including Goldstone (dashed) and ghost (dotted) internal lines, are
as follows.
\begin{center}
  \includegraphics{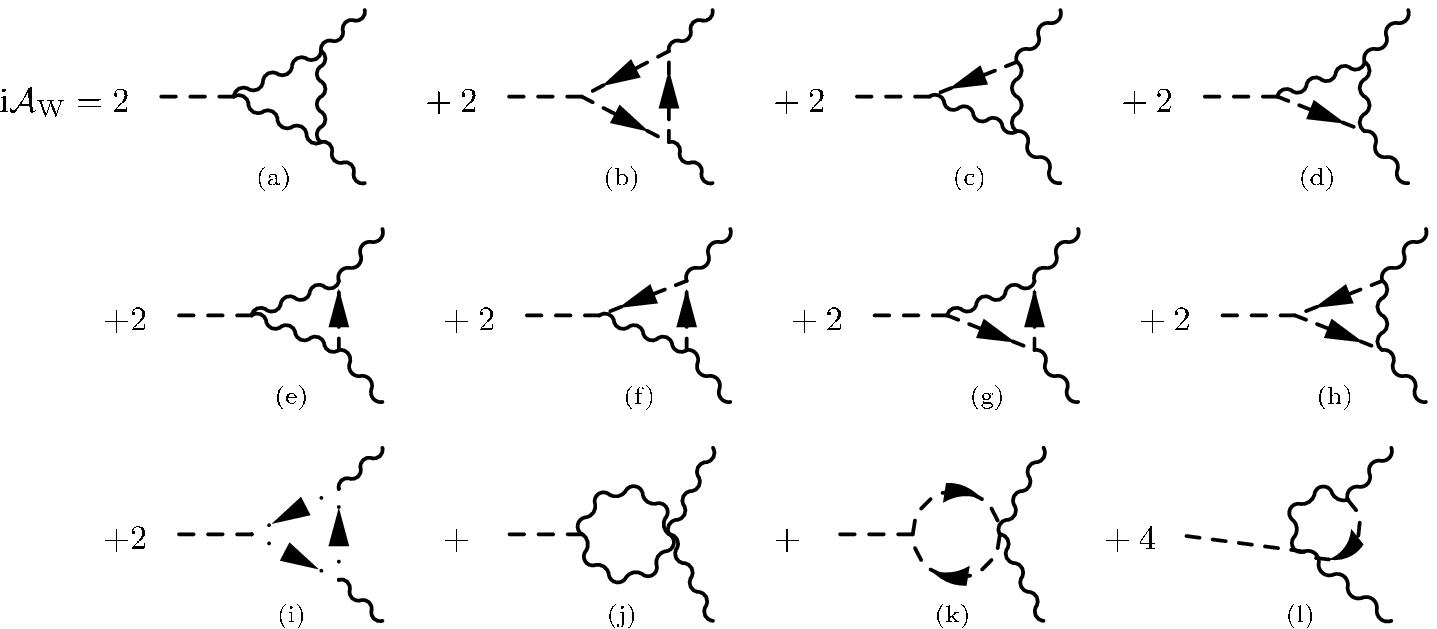}
\end{center}
In the following we calculate the general expression for each diagram in turn
and the corresponding SM and $n$th KK level expressions using the values for
the couplings in Sec.~\ref{app:feynmanrules}.

There are two $W$ diagrams (related by crossing the external photons):
\begin{align*}
\ii(\mathcal{A}_{a})_{\mu\nu} &= 2\times(\ii G_{WW})^{2}(\ii\lambda_{WW})(-\ii)^{3}\int_{k}\frac{1}{\mathcal{D}} g^{\rho\sigma}[-(k+p)_{\mu}g_{\sigma\lambda}+(k+p)_{\sigma}g_{\mu\lambda}-p_{\lambda}g_{\mu\sigma}+p_{\sigma}g_{\mu\lambda} \\
&\quad +k_{\lambda}g_{\mu\sigma}-k_{\mu}g_{\sigma\lambda}]g^{\lambda\kappa}g_{\kappa\alpha}g^{\alpha\beta}[-k_{\nu}g_{\beta\rho}+k_{\beta}g_{\nu\rho}-q_{\rho}g_{\nu\beta}+q_{\beta}g_{\nu\rho}+(k-q)_{\rho}g_{\nu\beta}-(k-q)_{\nu}g_{\beta\rho}] \\
&= 2G_{WW}^{2}\lambda_{WW} \frac{\ii\pi^{2}}{(2\pi)^{4}}[2g_{\mu\nu}g^{\rho\sigma}C_{\rho\sigma}+g_{\mu\nu}(p-q)^{\rho}C_{\rho}-\frac{5g_{\mu\nu}m_{h}^{2}}{2}C_{0}+10C_{\mu\nu}+p_{\nu}C_{\mu}-q_{\mu}C_{\nu} \\
&\quad\quad\quad\quad\quad\quad\quad\quad\quad\quad +4p_{\nu}q_{\mu}C_{0}] \\
&= \frac{\ii G_{WW}^{2} \lambda_{WW}}{8\pi^{2}}[2Dg_{\mu\nu} C_{00}-2m_{h}^{2}g_{\mu\nu}C_{12} - g_{\mu\nu}m_{h}^{2}C_{1} -  \frac{5g_{\mu\nu}m_{h}^{2}}{2}C_{0} + 10g_{\mu\nu}C_{00}- 10p_{\nu}q_{\mu}C_{12} \\
&\quad\quad\quad\quad\quad\quad\quad\quad\quad\quad-2p_{\nu}q_{\mu}C_{1}+ 4p_{\nu}q_{\mu}C_{0}] \\
&= \frac{\ii G_{WW}^{2}\lambda_{WW}}{8\pi^{2}}\bigg[
\left(D m^2 g_{\mu \nu } +3 m^2 g_{\mu \nu } -\frac{5}{2} m_h^2 g_{\mu\nu} -\frac{10 m^2 p_{\nu } q_{\mu }}{m_h^2}+4 p_{\nu } q_{\mu }\right)C_{0}\\
&\quad +\left(\frac{1}{2} D g_{\mu\nu}-\frac{2 p_{\nu } q_{\mu }}{m_h^2} +\frac{3}{2} g_{\mu\nu}\right) B_0(m_h^2,m^2)+\left(g_{\mu\nu}+\frac{2 p_{\nu } q_{\mu}}{m_h^2}\right)B_0(0,m^2)+\frac{D g_{\mu\nu}}{2}+\frac{3 g_{\mu\nu}}{2} \\
&\quad -\frac{5 p_{\nu } q_{\mu }}{m_h^2}\bigg] \\
&= \frac{\ii G_{WW}^{2}\lambda_{WW}}{8\pi^{2}}\bigg[
\left(4 m^2 g_{\mu \nu } +3 m^2 g_{\mu \nu } -\frac{5}{2} m_h^2 g_{\mu\nu} -\frac{10 m^2 p_{\nu } q_{\mu }}{m_h^2}+4 p_{\nu } q_{\mu }\right)C_{0}\\
&\quad +\left(2 g_{\mu \nu }-\frac{2 p_{\nu } q_{\mu }}{m_h^2} +\frac{3}{2} g_{\mu\nu}\right) B_0(m_h^2,m^2)+\left(g_{\mu\nu}+\frac{2 p_{\nu } q_{\mu}}{m_h^2}\right)B_0(0,m^2)+\frac{5 g_{\mu\nu}}{2} \\
&\quad -\frac{5 p_{\nu } q_{\mu }}{m_h^2}\bigg].
\end{align*}
Throughout the calculation we work in $D=4-\epsilon$ dimensions except for the last equality where we take the $\epsilon\to 0^+$ limit. Care must be taken in the case of the first $B_0$ function:
\[
\lim_{\epsilon\to 0^+} [D B_0(m_h^2,m^2)] = 4\lim_{\epsilon\to 0^+}[B_0(m_h^2,m^2)] - 2;
\]
this is the origin of the extra $-g_{\mu\nu}$ term in the last line.

The two Goldstone loop diagrams evaluate to
\begin{align*}
\ii(\mathcal{A}_{b})_{\mu\nu} &= 2\times(\ii G_{GG})^2 (\ii\lambda_{GG})(+\ii)^3
\int_k \frac{1}{\mathcal{D}}[-(k+p)-k]_{\mu}[-k-(k-q)]_{\nu} \\
&= -2G_{GG}^{2}\lambda_{GG}\frac{\ii\pi^{2}}{(2\pi)^{4}} 4C_{\mu\nu} \\
&= -\frac{\ii G_{GG}^{2}\lambda_{GG}}{2\pi^{2}} (C_{00}g_{\mu\nu} -
p_{\nu}q_{\mu}C_{12}) \\
&= -\frac{\ii G_{GG}^{2}\lambda_{GG}}{2\pi^{2}} \left[
\left(\frac{m^{2}g_{\mu\nu}}{2} -
\frac{m^{2}p_{\nu}q_{\mu}}{m_{h}^{2}}\right)C_{0} +
\frac{g_{\mu\nu}}{4}B_{0}(m_{h}^{2},m^2) + \left(\frac{g_{\mu\nu}}{4} -
\frac{p_{\nu}q_{\mu}}{2m_{h}^{2}}\right)\right].
\end{align*}

There are $3\times 2$ diagrams with two $W$s and one Goldstone. The first four
give\footnote{As in the case of $\tilde{\mathcal{A}}_a$, one must be careful when taking the $D\to4$ limit in the last equality.}

\begin{align*}
\ii(\mathcal{A}_{c})_{\mu\nu} = \ii(\mathcal{A}_d)_{\mu\nu} &= 2\times(-G_{WG})
(\lambda_{WG})(\ii G_{WW}) (-\ii)^{2}(+\ii) \int_{k}\frac{1}{\mathcal{D}}
g^{\rho\sigma}g_{\mu\sigma}  [-(p+q)-(k+p)]_{\lambda}\\
&\quad \times g^{\lambda\kappa} [-k_{\nu}g_{\kappa\rho} +k_{\kappa}g_{\nu\rho}
- q_{\rho}g_{\nu\kappa} +
q_{\kappa}g_{\nu\rho}+(k-q)_{\rho}g_{\nu\kappa}-(k-q)_{\nu}g_{\kappa\rho}] \\
&= \frac{\ii G_{WG}G_{WW}\lambda_{WG}}{8\pi^{2}} \\
&\quad\times [g_{\mu\nu} g^{\rho\sigma}C_{\rho\sigma} + 2g_{\mu\nu}(p+q)^{\rho}C_{\rho} + g_{\mu\nu}m_{h}^{2}C_{0} - C_{\mu\nu} +
2p_{\nu}C_{\mu} - 4q_{\mu}C_{\nu} - 4p_{\nu}q_{\mu}C_{0}] \\
&= \frac{\ii G_{WG}G_{WW}\lambda_{WG}}{8\pi^{2}} [(D-1)g_{\mu\nu}C_{00} -
(m_{h}^{2}g_{\mu\nu}-p_{\nu}q_{\mu})C_{12} + (m_{h}^{2}g_{\mu\nu} -
4p_{\nu}q_{\mu})C_{1} \\
&\quad \quad- (m_{h}^{2}g_{\mu\nu}+2p_{\nu}q_{\mu})C_{2} +
(m_{h}^{2}g_{\mu\nu}-4p_{\nu}q_{\mu})C_{0}] \\
&= \frac{\ii G_{WG}G_{WW}\lambda_{WG}}{8\pi^{2}}
\bigg[\left(\frac{(D-1) g_{\mu\nu}}{4} -
\frac{6p_{\nu}q_{\mu}}{m_{h}^{2}}\right) B_{0}(m_{h}^{2},m^2)
+ \frac{6p_{\nu}q_{\mu}}{m_{h}^{2}}B_{0}(0,m^2) \\
&\quad\quad\quad\quad +\left(\frac{(D-3)}{2} m^2 g_{\mu\nu} + m_{h}^{2}
g_{\mu\nu} + \frac{m^2}{m_{h}^{2}} p_{\nu} q_{\mu}-4
p_{\nu}q_{\mu}\right)C_{0}
+ \frac{(D-3)g_{\mu\nu}}{4} + \frac{p_{\nu}q_{\mu}}{2m_{h}^{2}} \bigg] \\
&= \frac{\ii G_{WG}G_{WW}\lambda_{WG}}{8\pi^{2}}
\bigg[\left(\frac{3}{4}g_{\mu\nu} - \frac{6p_{\nu}q_{\mu}}{m_{h}^{2}}\right)
B_{0}(m_{h}^{2},m^2)-\frac{g_{\mu\nu}}{2} +
\frac{6p_{\nu}q_{\mu}}{m_{h}^{2}}B_{0}(0,m^2) \\
& \quad\quad\quad \left(\frac{m^{2}}{2}g_{\mu\nu} + m_{h}^{2}g_{\mu\nu} +
\frac{m^{2}}{m_{h}^{2}}p_{\nu}q_{\mu} - 4p_{\nu}q_{\mu}\right)C_{0} +
\frac{g_{\mu\nu}}{4} + \frac{p_{\nu}q_{\mu}}{2m_{h}^{2}}\bigg].
\end{align*}

The second two yield
\begin{align*}
\ii(\mathcal{A}_{e})_{\mu\nu} &= 2\times(G_{WG})(-G_{WG})(\ii\lambda_{WW})(\ii)(-\ii)^2 \int_k\frac{1}{\mathcal{D}} \left(g_{\mu\rho} g^{\rho\sigma} g_{\sigma\lambda} g^{\lambda\kappa} g_{\kappa\nu}\right) \\
&= \frac{\ii \lambda_{WW}G_{Wf}^{2}g^{\mu\nu}}{8\pi^{2}}C_{0}.
\end{align*}

There are similarly $3\times 2$ diagrams involving one $W$ and two Goldstones. The first four evaluate to
\begin{align*}
\ii(\mathcal{A}_{f})_{\mu\nu} = \ii(\mathcal{A}_{g})_{\mu\nu} &= 2\times (\ii G_{GG})(\lambda_{WG})(-G_{WG}) (+\ii)^{2}(-\ii) \\
&\quad \int_{k}\frac{1}{\mathcal{D}}[-(k+p)-k]_{\mu} [-(p+q)-(k+p)]_{\rho}g^{\rho\sigma}g_{\nu\sigma} \\
&= 2G_{GG}G_{WG}\lambda_{WG}\int_{k}\frac{1}{\mathcal{D}} 2k_{\mu}(2p+q+k)_{\nu} \\
&= \frac{\ii G_{GG}G_{WG}\lambda_{WG}}{4\pi^{2}} (2p_{\nu}C_{\mu}+C_{\mu\nu}) \\
&= \frac{\ii G_{GG}G_{WG}\lambda_{WG}}{4\pi^{2}} (-2p_{\nu}q_{\mu}C_{1} + g_{\mu\nu}C_{00} - p_{\nu}q_{\mu} C_{12}) \\
&= \frac{\ii G_{GG}G_{WG}\lambda_{WG}}{4\pi^{2}} \bigg[
\left(\frac{m^{2}g_{\mu\nu}}{2} - \frac{m^{2} p_{\nu}q_{\mu}}{m_{h}^{2}}\right)C_{0}+
\left(\frac{g_{\mu\nu}}{4}-\frac{2p_{\nu}q_{\mu}}{m_{h}^{2}}\right)B_{0}(m_{h}^{2},m^2) \\
&\quad\quad\quad\quad\quad\quad\quad\quad\quad\quad + \frac{2p_{\nu}q_{\mu}}{m_{h}^{2}}B_{0}(0,m^2) + \left( \frac{g_{\mu\nu}}{4}- \frac{p_{\nu}q_{\mu}}{2m_{h}^{2}}\right)
\bigg]
\end{align*}
while the other two give
\begin{align*}
\ii(\mathcal{A}_{h})_{\mu\nu} &= 2\times(-G_{WG})(+G_{WG})(+\ii \lambda_{GG})(\ii)^{2}(-\ii)\int_k\frac{1}{\mathcal{D}} g^{\rho\sigma}g_{\mu\sigma} g_{\nu\rho} \\
&= \frac{\ii \lambda_{GG}G_{WG}^{2}g_{\mu\nu}}{8\pi^{2}} C_{0}.
\end{align*}

The last triangle diagrams are the two involving Faddeev-Popov ghosts:
\begin{align*}
  \ii(\mathcal{A}_{i})_{\mu\nu} &= -2\times(-\ii G_{\bar{c}c})^2 (+\ii \lambda_{\bar{c}c})(+\ii)^3 \int_k \frac{1}{\mathcal{D}} [-(k+p)]_\mu (-k)_\nu \\
  &= 2G_{\bar{c}c}^{2} \lambda_{\bar{c}c} \frac{\ii\pi^{2}}{(2\pi)^{4}} C_{\mu\nu} \\
  &= \frac{\ii G_{\bar{c}c}^{2}\lambda_{\bar{c}c}}{8\pi^2}\left(C_{00} g_{\mu\nu} - p_\nu q_\mu C_{12}\right) \\
  &= \frac{\ii G_{\bar{c}c}^{2}\lambda_{\bar{c}c}}{8\pi^2}\left[\left(\frac{m^2}{m_h^2}C_0 + \frac{1}{2m_h^2}\right)\left(\frac{m_h^2 g_{\mu\nu}}{2}-p_\nu q_\mu\right) + \frac{g_{\mu\nu}}{4}B_0(m_h^2,m^2)\right].
\end{align*}

There are six remaining (non-triangle) diagrams involving four-point vertices.
The $W$ diagram evaluates to
\begin{align*}
\ii(\mathcal{A}_{j})_{\mu\nu} &= (\ii H_{WW})(\ii
\lambda_{WW})(-\ii)^{2} \int_{k}\frac{g^{\rho\sigma}}{k^{2}-m^{2} +
\ii\epsilon} g_{\sigma\lambda} \frac{g^{\lambda\kappa}}{(k-p-q)^{2}-m^{2} +
\ii\epsilon}(2g_{\mu\nu}g_{\kappa\rho}
- g_{\mu\rho}g_{\nu\kappa} - g_{\mu\kappa}g_{\nu\rho})
\\
&= H_{WW}\lambda_{WW}2g_{\mu\nu}(D-1)
\frac{\ii\pi^{2}}{(2\pi)^{4}}B_{0}(m_{h}^{2},m^{2}) \\
&= \frac{\ii H_{WW} \lambda_{WW}(D-1)g_{\mu\nu}}{8\pi^{2}}
B_{0}(m_{h}^{2},m^{2}) \\
&= \frac{\ii H_{WW}\lambda_{WW}g_{\mu\nu}}{8\pi^{2}}[3B_{0}(m_{h}^{2},m^{2}) -
2],
\end{align*}
and the Goldstone diagram evaluates to
\begin{align*}
\ii(\mathcal{A}_{k})_{\mu\nu} &= (\ii H_{GG})(\ii \lambda_{GG})(\ii)^{2}g_{\mu\nu}
\int_{k}\frac{1}{k^2-m^2 + \ii\epsilon}\frac{1}{(k-p-q)^2-m^2+ \ii\epsilon} \\
&= \frac{\ii H_{GG} \lambda_{GG} g_{\mu\nu}}{16\pi^{2}}B_{0}(m_{h}^{2},m^{2}).
\end{align*}
The last four diagrams are related to diagram (l) (shown above) by swapping external momenta and changing the direction of internal particle number flow:
\begin{align*}
\ii(\mathcal{A}_{l})_{\mu\nu} &= 4\times (-G_{WG})(+I_{WG})(-\ii)(+\ii)
\int_{k}\frac{g^{\rho\sigma} g_{\mu\sigma}g_{\nu\rho}}{(k^{2}-m^{2} +
\ii\epsilon)[(k+p)^{2}-m^{2} +\ii\epsilon]} \\
&= -\frac{\ii G_{WG}I_{WG} g_{\mu\nu}}{4\pi^{2}} B_{0}(0,m^2).
\end{align*}

\subsection{Summing the diagrams}
We now need to sum the diagrams to find the SM and $n$th KK level contributions. We have checked the following expressions in \texttt{Mathematica}.

Putting in the SM values for masses and couplings gives the following. The Goldstone and ghost have the same mass as the $W$ boson. (The same applies for the higher KK modes.)
\begin{align*}
\ii(\mathcal{A}_{W}^{\text{SM}})_{\mu\nu} &= \frac{3 i e^2 g m_{h}^2 m_{W} g_{\mu \nu } C_{0}}{8 \pi ^2}+\frac{3 i e^2 g m_{W}^3 g_{\mu \nu } C_{0}}{4 \pi ^2}+\frac{3 i e^2 g m_{W} p^{\nu } q^{\mu } C_{0}}{4 \pi ^2}-\frac{3 i e^2 g m_{W}^3 p_{\nu } q_{\mu } C_{0}}{2 \pi ^2 m_{h}^2} \\
&+\frac{i e^2 g m_{h}^2 g_{\mu \nu }}{16 \pi ^2 m_{W}}+\frac{3 i e^2 g m_{W} g_{\mu \nu }}{8 \pi ^2}-\frac{3 i e^2 g m_{W} p_{\nu } q_{\mu }}{4 \pi ^2 m_{h}^2}-\frac{i e^2 g p_{\nu } q_{\mu }}{8 \pi ^2 m_{W}}.
\end{align*}
Factoring out the Lorentz part and writing things in terms of the dimensionless function $f_V$, defined in \eqref{eq:fV}, gives
\[
\tilde{\mathcal{A}}_W = -\frac{e^{2} g}{16\pi^{2}m_{W}}f_V(m_W).
\]

The sum of the diagrams at the $n$th KK level is
\[
\ii(\mathcal{A}_{W}^{(n)})_{\mu\nu} = \frac{\ii e^{2} g m_{W}}{8\pi^{2}m_{h}^{2}m_{W,n}^{2}}\left[m_{h}^{2} + 6m_{W,n}^{2}+\left(12 m_{W,n}^{4} - 6 m_{h}^{2}m_{W,n}^{2}\right)C_{0}\right]\left(\frac{m_{h}^{2} g_{\mu\nu}}{2}-p_{\nu}q_{\mu}\right),
\]
so
\[
\tilde{\mathcal{A}}_{W}^{(n)} = -\frac{e^{2}g}{16\pi^{2}m_{W}}\left(\frac{m_W}{m_{W,n}}\right)^2 f_{V}(m_{W,n}).
\]

\subsection{Scalar contribution}
For KK number $n\ge 1$ there exist charged scalar particles $a^{\pm}_n$
not seen at the SM level. At tree level these have the same masses as their
$W^\pm_n$ counterparts but loop corrections split this degeneracy.

There are three allowed diagrams at each KK level contributing to the
$h\to\gamma\gamma$ amplitude that involve a charged scalar (two of them are
numerically equal and are related by swapping the photon momenta):

\[
\ii(\mathcal{A}_{a^\pm}^{(n)})_{\mu\nu} = 2\times\; \begin{array}{c}\includegraphics[scale=0.75]{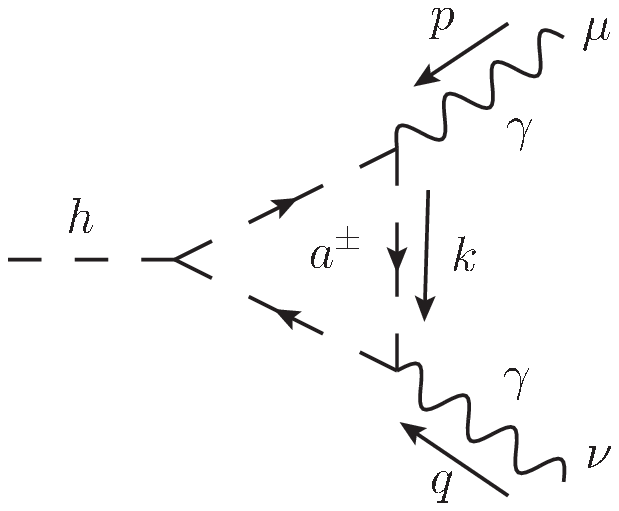}\end{array}
+
\begin{array}{c}
	\includegraphics[scale=0.75]{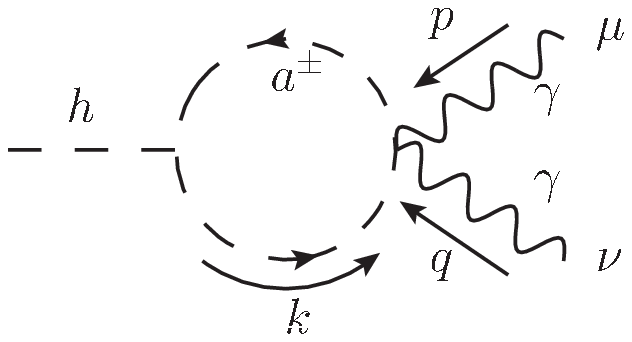}
\end{array}
\]
These diagrams are exactly the same as the similar Goldstone diagrams $\mathcal{A}_b$ and $\mathcal{A}_k$ evaluated in the previous section, but with a different particle mass and different couplings. Using the couplings from Sec.~\ref{app:feynmanrules}, we get that
\begin{align*}
\ii (\mathcal{A}_{a^\pm}^{(n)})_{\mu\nu} &= -\frac{2\ii e^{2}\lambda_{aa}}{4\pi^{2}} \left[ \left(\frac{m_{a,n}^{2}g_{\mu\nu}}{2} - \frac{m_{a,n}^{2}p_{\nu}q_{\mu}}{m_{h}^{2}}\right)C_{0} + \frac{g_{\mu\nu}}{4} - \frac{p_{\nu}q_{\mu}}{2m_{h}^{2}}\right] \\
&= \frac{\ii e^{2} g}{4\pi^{2} m_{h}^{2} m_{W}} \left[2\frac{m_{a,n}^{2}}{m_{W,n}^{2}}m_{W}^{2} + m_{h}^{2}\left(1-\frac{m_{W}^{2}}{m_{W,n}^{2}}\right)\right]
\left(\frac{1}{2}+m_{a,n}^{2}C_{0}\right)\left(\frac{m_{h}^{2}g_{\mu\nu}}{2}-p_{\nu}q_{\mu}\right),
\end{align*}
so
\[
\tilde{\mathcal{A}}_{a^\pm}^{(n)} = -\frac{e^{2} g}{16\pi^{2} m_{W}} f_S(m_{a,n},m_{W,n}),
\]
where $f_S(m_{a,n},m_{W,n})$ is defined in \eqref{eq:fS}.

\section{P-V functions and conventions}
\label{app:pv}

\subsection{Three-point PV function}

The three-point scalar Passarino-Veltman function is frequently encountered
when evaluating triangle diagrams. It is defined by
\[
\frac{\ii\pi^{2}}{(2\pi)^{4}}C_{0}(p^{2},(p+q)^{2},q^{2},m_{0}^{2},m_{1}^{2},m_{2}^{2}) \equiv\mu^{4-D} \int\frac{\d^{D}k}{(2\pi)^{D}} \{(k^{2}-m_{0}^{2})[(k+p)^{2}-m_{1}^{2}][(k-q)^{2}-m_{2}^{2}]\}^{-1}.
\]
We encounter this integral exclusively in the special case that the internal masses are equal:
\[
\frac{\ii\pi^{2}}{(2\pi)^{4}}C_{0} \equiv \mu^{4-D} \int\frac{\d^{D}k}{(2\pi)^{D}} \frac{1}{\D},
\]
where $\D$ is the denominator from the general expression with $m_{0}=m_{1}=m_{2}\equiv m$:
\[
  \D = (k^2 - m_0^2)[(k+p)^2 - m^2][(k-q)^2 - m^2].
\]
This integral can be evaluated \cite{tHooft:1978xw} to give
\begin{equation}
C_{0}(m) = 
\begin{cases}
	-\frac{2}{m_h^2}\left[\arcsin\left(\frac{m_{h}}{2m}\right)\right]^{2} & m^{2} \ge m_{h}^{2}/4 \\
	\frac{1}{2m_h^2}\left[\ln\left(\frac{1+\sqrt{1-4m^{2}/m_{h}^{2}}}{1-\sqrt{1-4m^{2}/m_{h}^{2}}}\right) - i\pi\right]^{2} & m^{2} < m_{h}^{2}/4. 
\end{cases}
\label{eq:c0dim}
\end{equation}

It is convenient to define a dimensionless version of this expression:
\[
c_{0}(m)=-\frac{m_{H}^{2}}{2}C_{0}(m),
\]
(where the normalisation matches the \texttt{fiRe} and \texttt{fiIm} functions
defined in the \texttt{SLHAplus} library for
\texttt{CalcHEP}/\texttt{MicrOMEGAS} \cite{Belanger:2010st}).

\subsection{Two-point PV function}

We also frequently come across the scalar two-point PV function:
\[
\frac{\ii\pi^{2}}{(2\pi)^{4}}B_{0}(p^{2},m^{2})\equiv\mu^{4-D}\int\frac{\d^{D} k}{(2\pi)^{D}}\{(k^{2}-m^{2})[(k+p)^{2}-m^{2}]\}^{-1}.
\]
We encounter the following two spacial cases
\begin{equation}
B_{0}(m_{h}^{2},m^{2}) =\frac{1}{\bar{\epsilon}}-\ln\frac{m^{2}}{\mu^{2}} + 2 -\sqrt{1-4m^2/m_h^2}\sqrt{2m_{h}^{2}C_{0}(m)}
\label{eq:b0mh2}
\end{equation}
and
\begin{equation}
B_{0}(0,m^{2}) = \frac{1}{\bar{\epsilon}}-\ln\frac{m^{2}}{\mu^{2}},
\label{eq:b00}
\end{equation}
with $D=4-\epsilon$ and
\begin{equation}
\frac{1}{\bar{\epsilon}} =\frac{2}{\epsilon} -\gamma_{E}-\ln\pi,
\label{eq:epsilon}
\end{equation}
$\gamma_E \approx 0.57721$ being the Euler-Mascheroni constant.

\subsection{PV Reduction}

More generally, we come across three-point momentum integrals with more
complex Lorentz structure that can be written generically as
\[
 \frac{\ii\pi^{2}}{(2\pi)^{4}}C^{\rho\sigma\ldots\kappa}\equiv\mu^{4-D}\int\frac{\d^{D}k}{(2\pi)^{D}} \frac{k^{\rho}k^{\sigma}\cdots k^{\kappa}}{\D}.
\]
These can be simplified through \emph{Passarino-Veltman reduction} to
expressions involving the scalar three-point and two-point PV functions
defined in \eqref{eq:c0dim}, \eqref{eq:b0mh2} and \eqref{eq:b00}; for
on-shell
external momenta ($p^{\mu}$ and $q^{\nu}$) we get
\begin{equation}
\begin{aligned}
C^{\mu\nu} &= g^{\mu\nu}C_{00} - p^{\nu}q^{\mu}C_{12} \\
g_{\rho\sigma}C^{\rho\sigma} &= DC_{00}-m_{h}^{2}C_{12} \\
C^{\mu} &= -q^{\mu}C_{2} \\
C^{\nu} &= p^{\nu}C_{1}.
\end{aligned}
\label{eq:PVtensred}
\end{equation}

The coefficient functions expand further to
\begin{equation}
\begin{aligned}
C_{00} &= \frac{1}{2} m^{2} C_0+\frac{1}{4} B_0(m_{h}^{2},m^{2})+\frac{1}{4} \\
C_{12} &= \frac{m^{2}}{m_{h}^{2}}C_0+\frac{1}{2 m_{h}^{2}} \\
C_{1}=C_{2} &= \frac{B_0\left(m_{h}^2,m^2\right)}{m_{h}^2}-\frac{B_0\left(0,m^2\right)}{m_{h}^2}.
\end{aligned}
\label{eq:PVred}
\end{equation}
We used the \texttt{PaVeReduce} function in the \texttt{FeynCalc} package for \texttt{Mathematica} to check this.

Ultimately one must take the $D\to 4$ limit. Particular care must be taken in the case of 
$C_{00}$:
\[
\lim_{D\to 4}DC_{00} = 4C_{00} + \lim_{\epsilon\to 0}\frac{\epsilon}{4\bar{\epsilon}}=4C_{00} +\frac{1}{2}.
\]